\PassOptionsToPackage{table}{xcolor}
\documentclass[a4paper,onecolumn,11pt]{quantumarticle}
\pdfoutput=1
\usepackage[utf8]{inputenc}
\usepackage[english]{babel}
\usepackage[T1]{fontenc}
\input{glyphtounicode}\pdfgentounicode=1

\usepackage{amsmath,amssymb,amsthm}
\newtheorem{proposition}{Proposition}
\usepackage{graphicx}
\usepackage{booktabs}
\usepackage{array}
\usepackage{float}
\usepackage{needspace}
\newcolumntype{L}[1]{>{\raggedright\arraybackslash}p{#1}}
\usepackage{caption}
\usepackage{placeins}
\usepackage{enumitem}
\usepackage{hyperref}
\usepackage{xcolor}
\definecolor{ours}{HTML}{FDEBDD}
\definecolor{stripe}{HTML}{F2F3F5}
\graphicspath{{figures/}}
\setlist{itemsep=2pt,topsep=4pt}

\newcommand{\ket}[1]{| #1 \rangle}
\newcommand{\braket}[2]{\langle #1 | #2 \rangle}

\begin{document}

\title{Reading Why: Born-Measured Attention with Exact Word Evidence on Light and Qubits}

\author{Teerawat Tongloy}
\orcid{0009-0001-0670-3124}
\email{teerawat.to@kmitl.ac.th}
\affiliation{School of Integrated Innovative Technology, King Mongkut's Institute of Technology Ladkrabang, Bangkok,
Thailand}
\author{Jiraporn Pongsopa}
\orcid{0009-0004-7410-9145}
\email{jiraporn\_pon@dusit.ac.th}
\affiliation{Faculty of Science and Technology, Suan Dusit University, Bangkok, Thailand}
\author{Rangsan Jomtarak}
\orcid{0009-0009-5417-5571}
\email{rangsan\_jom@dusit.ac.th}
\affiliation{Faculty of Science and Technology, Suan Dusit University, Bangkok, Thailand}

\begin{abstract}
When a language model answers, we can usually only estimate why. We build a transformer in which every nonlinearity
is a quadratic detection---an interference, a balanced reading or a camera---and the output is a Born probability:
attention $|\braket{q}{k}|^2$, the form of the Born rule, normalised per row, a measured product for the
feed-forward nonlinearity, and a Born readout; between detections, classical control rescales and re-encodes.
With its denominators held at the input, each class probability is a polynomial in the word
amplitudes, so its integrated-gradient evidence is computed by a finite quadrature with no quadrature error: the
words' contributions sum to the class probability above its no-information share to $10^{-13}$. On e-SNLI
natural language inference it reaches $77.45 \pm 0.91\,\%$
accuracy, level with a same-size softmax transformer ($76.62 \pm 0.36\,\%$) and an order-free bag of embeddings
($77.79\,\%$). Exactness buys the sum, not the best ranking. Deleting the word its evidence ranks first changes the
answer in 44\,\% of pairs, of the 53\,\% some deletion changes, against 31--32\,\% for the softmax model's
integrated gradients; the model's own gradient $\times$ input does as well (43\,\%), as does a bag of embeddings'
(43\,\%, seed 0), and leave-one-out does better. Holding the denominators helps the low end: the five words ranked last
change the answer in 1\,\%, against 6\,\% free. The same weights compile exactly to simulated optoelectronics, and
their attention to qubits. On ibm\_kingston, reading only the attention, the device gave the model's answers in all
three Hadamard-test runs (the two shortest pairs and one without its deciding word) and both compute--uncompute runs
(one circuit per overlap); uniform attention answers otherwise on two of the three inputs. With the device's row sums the
evidence correlated with the exact one at 0.997 and 0.962 by Hadamard tests; with compute--uncompute, 0.971 and 0.48.
\end{abstract}

\section{Introduction}
\label{sec:intro}

Why a language model gives the answer it gives can usually only be estimated. Explanations of neural networks are mostly computed after training and approximate what the model did, and
for language models the most visible quantity, the attention weights, need not explain the output
\cite{jain2019attention,wiegreffe2019attention}. A model that is right for the wrong reasons looks, from its accuracy,
the same as one that is right for the right ones.

In earlier work, \emph{Seeing Why}~\cite{seeingwhy}, a picture classifier built from a Born readout gave evidence maps
that need no approximation, and that could be measured on the hardware that classifies. Pictures, however, are read
pixel by pixel; in a sentence, words matter through each other. Here we ask whether the same can be had for a model
that relates words: a transformer.

Attention needs two things: an overlap between a query and a key, and a weighted sum. Both are native to light and to
qubits: an overlap is an interference, a weighted sum a coherent combination. What is not native is the softmax. We
replace every nonlinearity of a transformer---softmax, LayerNorm, ReLU---by quadratic detections, interferences,
balanced interferometric readings and a camera, whose squared moduli give the rule Born gave a century ago~\cite{born1926}; the
attention weight takes the form of the Born rule, a squared overlap over its row sum. The model is then a ratio of polynomials in the word amplitudes, and its
evidence follows exactly from integrated gradients~\cite{sundararajan2017ig} with the denominators held.

Because the evidence is exact and part of the model, the model can be checked as well as scored: we report, for every
model, accuracy beside the completeness of its evidence, a deletion test and agreement with human highlights.

Where the exactness comes from should be said plainly. It comes from the model being a polynomial once its
denominators are held, and a classical polynomial network without any measurement constraint has it as well: an
unconstrained polynomial model of the same size, whose evidence is exact too but which is not built from measurements
(it differs also in its real parameters, readout and loss), is one and a half points more accurate (Section~\ref{sec:esnli}).
What the Born rule adds is that the polynomial's numerators and held denominators are formed from detected readings
(the camera's total detected directly, $\sum_t a_t$ set at the modulator):
the same exact evidence then comes with a model that light and qubits can run, and the questions become physical
ones---which measurements, how many shots, what loss of visibility the evidence tolerates.

For quantum information the work offers three things beyond the model. The row sums the evidence holds become
measured quantities whose error can be budgeted: a loss of visibility enters each attention weight squared and cancels
within a row, but not in those row sums, so we give the shots, calibration depth and per-row accuracy that a word's
evidence needs, and test the first two at one setting on a device (Section~\ref{sec:device}). Attention with no
measurement inside is one post-selected circuit whose success probability has a closed form
(Section~\ref{sec:coherent}). And since the attribution is a finite computation, the explanation can be computed with
row sums measured on the same hardware, for the probability those row sums give.

Light and qubits then pull the design in two directions. On light a measurement is cheap---a detector---and a field can be split into copies,
but classical fields cannot hold the copies together in a tensor product; on qubits they can, but every measurement costs shots. So we build
two models from the same idea: the \emph{Born Transformer}, which measures inside every block and suits light,
and a \emph{coherent attention layer}, which measures only at the end and suits qubits. Both keep exact evidence with their denominators held, and the attention of the first, read by small circuits, runs on a quantum computer today.

\begin{figure}[!htb]
  \centering
  \includegraphics[width=\linewidth]{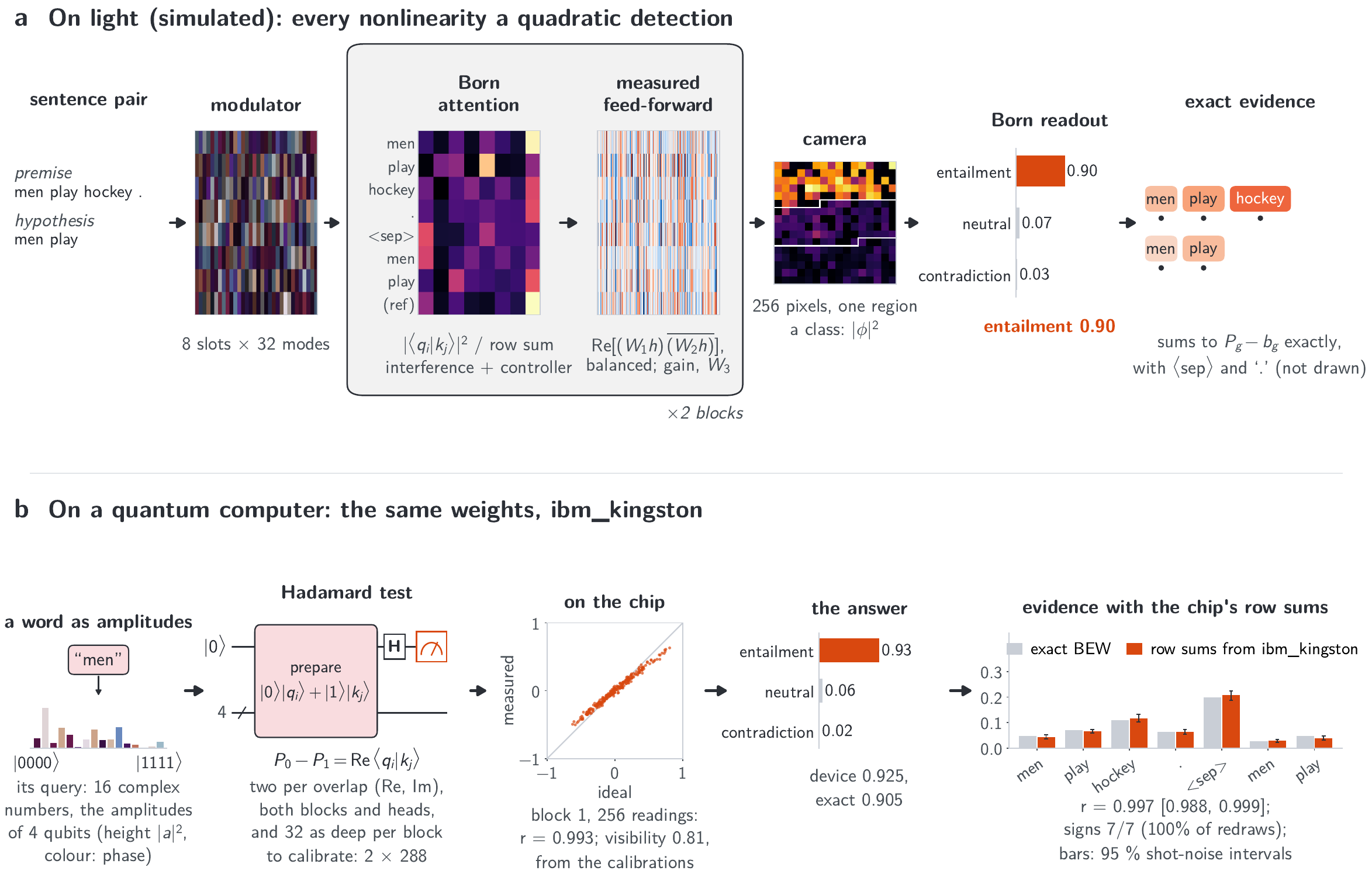}
  \caption{One test pair, ``men play hockey . / men play'' (entailment; test pair 3765), through the trained model.
  (a) On light (simulated): the modulator's fields, Born attention and the measured feed-forward in two blocks, the
  camera's class regions and each word's exact evidence (BEW; red for the answer, blue against; dots: the annotators'
  words). (b) On ibm\_kingston, only the attention overlaps read on the device: the Hadamard test, the readings
  against their ideal values, the answer, and each word's evidence with the device's row sums against the exact one
  (bars: 95\,\% shot-noise intervals; Section~\ref{sec:device}).}
  \label{fig:overview}
\end{figure}

Fig.~\ref{fig:overview} follows one test pair through the model, on light and on a quantum computer.
Our contributions:
\begin{enumerate}
\item \emph{A transformer whose every nonlinearity is a quadratic detection} (Section~\ref{sec:model}, Fig.~\ref{fig:compare3}):
  attention $|\braket{q}{k}|^2$, in the form of the Born rule, normalised over each row, a measured product in place of the feed-forward
  nonlinearity, and a Born readout, so that each class probability is a ratio of polynomials in the word amplitudes.
\item \emph{Born evidence per word} (BEW, Section~\ref{sec:evidence}): the word-level counterpart of the Born
  evidence maps (BEM) of Seeing Why~\cite{seeingwhy}. BEW extends BEM from a quadratic form to a deep, attention-based
  model: still exact with the denominators held---the words' evidence sums to the class probability above its no-information share---now by
  quadrature rather than in closed form. A degree bound and Gauss--Legendre quadrature also make the attributions of a
  polynomial graph network's pre-softmax scores exact~\cite{feng2026apex}; here the denominators are held, so that the
  class probability itself is decomposed at the input, and the polynomial is the one the measurements compute.
\item \emph{Natural language inference, checked beyond accuracy} (Section~\ref{sec:esnli}): on e-SNLI the model is
  level with a softmax transformer of the same size and with an order-free bag of embeddings, so accuracy does not
  set them apart; deleting the word its evidence ranks first changes its answer
  more often than with the softmax model's integrated gradients---a gain of the model, matched by its own free
  gradient $\times$ input and by a bag of embeddings' and exceeded by leave-one-out, while holding the denominators
  improves the words ranked last and the agreement---and its top words are closer than that model's evidence to those annotators highlighted, though rankings that need no model, by word position, are closer still; its attention weights, as close to the
annotators, account for the answer less well.
\item \emph{One set of weights, two realisations} (Section~\ref{sec:compile}): the trained weights compile exactly to optics (simulated, with shot noise and finite modulator precision), where an
  independent mesh decomposition confirms the build; on qubits the attention compiles to Hadamard tests, or to compute--uncompute circuits that read each Born weight
  $|\braket{k_j}{q_i}|^2$ as a probability (the other stages have circuit forms we did not build), and both
  run on hardware.
\item \emph{On a quantum computer, and what it costs} (Section~\ref{sec:device}): on ibm\_kingston, with the attention
  overlaps read by the device and the rest computed classically, the answer survives the device on the two shortest
  test pairs and follows the model when the one word that decides one of them is deleted (two of the three runs
  where uniform attention would answer otherwise, though random overlaps would pass them about a quarter and two fifths
  of the time), also with compute--uncompute circuits, with half the overlap circuits and
  fewer gates but larger errors; with Hadamard tests and calibration circuits as
  deep as the overlaps, BEW computed with the device's attention row sums follows the exact evidence on one pair closely
  and on the other within shot noise and a modest per-circuit error (it explains $\tilde P_g$, exact numerators over the measured row
  sums), a consistency check of the calibrated row sums, while with compute--uncompute it fails on one of its two
  runs; an analysis of shots, visibility and calibration says when it holds, and a proposition gives the shot-noise
  floor of the evidence, pair by pair from the noiseless model, and how it grows with the evidence's cancellation;
  on the device the shots were not the limit: with compute--uncompute, errors per circuit beyond shot noise set it.
\item \emph{A separate coherent attention layer, trained on its own} (Section~\ref{sec:coherent}): with no measurement until the end, attention is one post-selected circuit equal to its model in simulation (pass rate $10^{-5}$--$10^{-4}$ on the seven-token pairs, about $3\times10^{-6}$ on long ones); with
  brickwall maps it needs 852 CNOTs on 23 qubits, and at 67\,\% accuracy, little above a linear bag of words and well below a bag of embeddings, keeps high deletion scores; its
  post-selection has a closed form and a simulated scaling.
\end{enumerate}

\section{Model}
\label{sec:model}

A transformer has three kinds of nonlinearity: the softmax of attention, LayerNorm and the ReLU (or GELU) of the
feed-forward layer. None of them is native to light or qubits. A quadratic detection is: an interference, a balanced
interferometric reading or a camera returns a product of two amplitudes, and the Born rule turns the squared modulus of an amplitude
into a probability. We build every nonlinearity from such detections; between them a classical controller squares, divides and rescales
the measured numbers and writes them back as fresh light, and the output is a Born probability.

\subsection{Every nonlinearity a quadratic detection}

\paragraph{Encoding.} A sentence pair, premise and hypothesis joined by a separator token, becomes one state per word.
Word $t$ (of $n$ present) gets the complex field
\begin{equation}
  x_t = a_t\, e_t, \qquad e_t = \bigl(E_{w_t} + P_t\bigr) \odot \bigl(e^{i\omega_k t}\bigr)_{k=1}^{32},
  \qquad a_t = 1/\sqrt{n},
  \label{eq:encode}
\end{equation}
with a learned embedding $E_w \in \mathbb{C}^{32}$ (32 modes of light, or five qubits), a learned position field
$P_t$, and a phase $e^{i\omega_k t}$ on mode $k$ with learned frequencies $\omega_k$ (as in complex word embeddings~\cite{wang2020complex}; the form of a rotary position code~\cite{su2024roformer}, applied here to the embedding as a diagonal phase). The phase alone did not let attention find words by position in our synthetic tests, so
the field is added as well. A reference slot carries the sentence mean, $g\,\bar m$ with $\bar m = \sum_t a_t e_t / \sum_t a_t$ and a learned gain $g$, and every
word keeps its contrast $x_t - A_t \bar m$, with $A_t = a_t$ on a plain pass. When explaining, the sum $\sum_t a_t$
and the coefficient $A_t$ are both held at their values for the input, so every field---each word's contrast and
the reference slot---is linear in the amplitudes $a_t$, which are what the evidence is about, and vanishes when they
are all zero.

\paragraph{Born attention.} Each of two heads maps every slot to a query, key and value of 16 modes,
$q_i = Q h_i$, $k_j = K h_j$, $v_j = V h_j$, and weighs the values by
\begin{equation}
  w_{ij} = \frac{|\braket{q_i}{k_j}|^2}{\sum_{j'} |\braket{q_i}{k_{j'}}|^2}, \qquad
  h_i \leftarrow h_i + O \sum_j w_{ij} v_j .
  \label{eq:attention}
\end{equation}
The sum runs over the words and the reference slot. The overlap $\braket{q_i}{k_j}$ is an interference---balanced
detection of two light fields, or a Hadamard test on qubits (Section~\ref{sec:compile}), each returning the real or
the imaginary part---and $w_{ij}$ has the form of the Born rule: the squared overlap of the query's state with the
key's, normalised over the row. The controller forms the squared modulus and the row sum from the detected readings. Born-rule attention of this form was introduced
by CNet~\cite{crasmaru2026cnet} for causal character-level language modelling; here it is bidirectional, explained
exactly and compiled to hardware. Normalised polynomials of the query--key product also stand in for the softmax in
efficient transformers~\cite{kacham2023polysketch,nauen2024taylorshift}, as do learned feature maps that mimic its
weights~\cite{zhang2024hedgehog}. Squaring the weight again,
$|\braket{q_i}{k_j}|^4$ over its row sum, sharpens it on synthetic tasks (Appendix~\ref{app:details}), but on the
task of this paper it is level on accuracy and deletion, and 0.01 lower in agreement, at 3.2 times the degree (Table~\ref{tab:esnli}), so we keep
the Born rule's form. A uniform loss of visibility scales every overlap of a row alike and cancels in $w_{ij}$.

\paragraph{Measured feed-forward.} In place of linear--ReLU--linear, the state is read as a product of two linear
images, and the reading is written back:
\begin{equation}
  r_c = \mathrm{Re}\bigl[(W_1 h)_c \, \overline{(W_2 h)_c}\bigr], \qquad h \leftarrow h + W_3\, r ,
  \label{eq:mlp}
\end{equation}
with 170 channels $c$. Each $r_c$ is a balanced interferometric reading of two fields---the two beat on a beam splitter and the difference of its two detectors is read, with no separate local oscillator (on qubits, a Hadamard test); after it a
classical gain and a fresh modulator encode $W_3 r$. The residual additions are passive combiners. There is no
LayerNorm: attention~\eqref{eq:attention} and the readout below are ratios and do not see a common scale of the
state, but the feed-forward reading is quadratic and does, so the model is not scale-free: scaling every amplitude by
1.5 changes the answer on 9.5\,\% of 200 test pairs (seed 0; 5.5 and 7.5\,\% for seeds 1 and 2;
\texttt{esnli\_baselines.py}). This matters for deletion
(Section~\ref{sec:esnli}), which renormalises the other words' amplitudes.

\paragraph{Born readout.} After two blocks the slots are combined coherently and read by a camera of $N = 256$ pixels,
split into one region $S_g$ per class:
\begin{equation}
  \phi = R \sum_t h_t, \qquad P_g = \frac{\sum_{b \in S_g} |\phi_b|^2}{\sum_b |\phi_b|^2}, \qquad
  b_g = \frac{|S_g|}{N},
  \label{eq:readout}
\end{equation}
where $b_g$ is the share a class gets with no information, set by how the camera's pixels are allotted to the
classes (86, 85 and 85 of 256 here), so the evidence explains the answer above a share the design fixes. Each detected photon lands in region $S_g$ with probability
$P_g$; with coherent light the pixel counts are independent Poisson counts, so a camera estimates $P_g$ from many
photons, while one photon gives only a sample. The model is trained on the cross-entropy of $P_g$ itself.
(The Born headroom readout of Seeing Why~\cite{seeingwhy}, a fixed-scale softmax of each class's headroom above its
no-information share, was fragile on sequences: on a synthetic task of finding a value by its key a real-valued variant trained to
19.7\,\% where $P_g$ itself reached 100\,\%; with $P_g$ itself the explained quantity is also the answer.) We call
the model the \emph{Born Transformer}; it has 1.53M parameters, 98k of them outside the embedding.

\begin{figure}[tp]
  \centering
  \includegraphics[width=\linewidth]{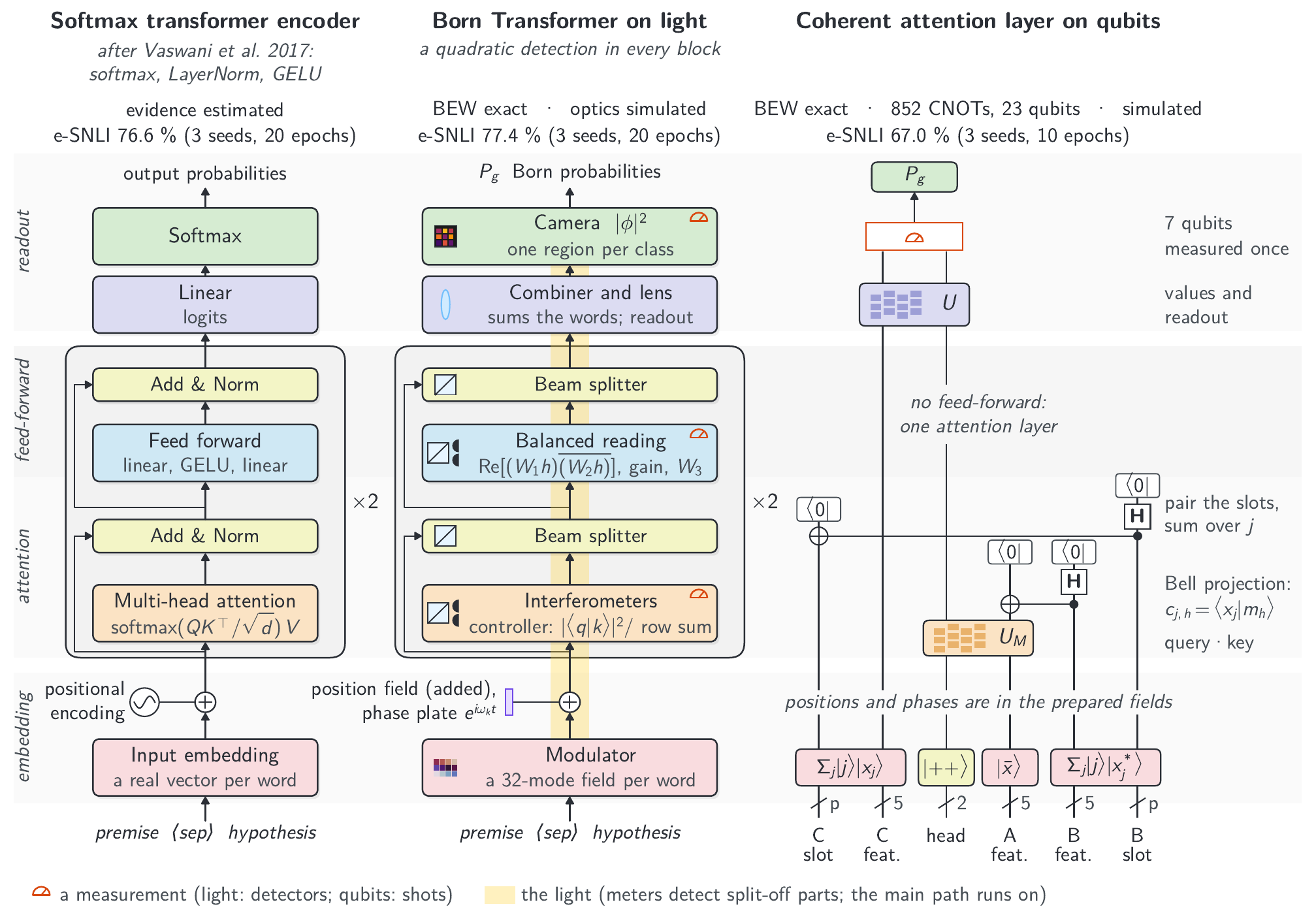}
  \caption{Three models step by step, bottom to top, in the box colours of the original transformer figure: a
  softmax transformer encoder after Vaswani et al.~\cite{vaswani2017}, the Born Transformer drawn as optics, and the
  coherent attention layer drawn as its circuit, wires running upwards (one circuit, one measurement). Bands hold the
  same stage; detector symbols mark the measurements.}
  \label{fig:compare3}
\end{figure}

Fig.~\ref{fig:compare3} sets the model against the original transformer~\cite{vaswani2017} step by step, beside the
coherent attention layer of Section~\ref{sec:coherent}: every nonlinearity of the original---softmax, LayerNorm,
ReLU (GELU in our baseline)---becomes a quadratic detection, or, in the coherent layer, the one measurement at the end. On light the words are
fields through a modulator, read by interferometers, balanced detectors and a camera. The same weights also compile
to qubits, every reading a Hadamard test (Section~\ref{sec:compile}); that realisation, whose attention ran on
ibm\_kingston, is drawn in Section~\ref{sec:device}. The coherent layer is a separate, smaller model, simulated
only. The encoder is drawn in the layout of the original; our softmax baseline places LayerNorm before each
sublayer, averages the words before the linear head and has two blocks, as the Born Transformer.

\subsection{Exact evidence: Born evidence per word}
\label{sec:evidence}

Every step above is a product, a sum or a squared modulus, and every division has the form of
Eqs.~\eqref{eq:attention} and \eqref{eq:readout}. So, with the denominators held at their values for the input (the
row sums of attention, the sum $\sum_t a_t$ in the sentence mean, the coefficients $A_t$ of the words' contrasts, and the total intensity
$Z = \sum_b|\phi_b|^2$), $F_g(a) = N_g(a)/Z$ with $N_g(a) = \sum_{b \in S_g}|\phi_b(a)|^2 - b_g \sum_b|\phi_b(a)|^2$ is a
polynomial in the amplitudes $a = (a_1, \dots, a_n)$; at the input it equals $P_g - b_g$, and $F_g(0) = 0$ since every
term of $N_g$ carries an amplitude. Its degree is bounded
step by step: the fields have degree 1, an overlap $\braket{q_i}{k_j}$ degree 2, $|\braket{q_i}{k_j}|^2$ degree 4, the
attention update (times a value) degree 5, the measured feed-forward doubles it to 10 per block, two blocks give
$10^2 = 100$, and the Born readout's squared modulus 200. The evidence of word $t$ for class $g$ is
integrated gradients from the zero baseline~\cite{sundararajan2017ig}, and it adds up to the answer:
\begin{equation}
  e_g(t) = a_t \int_0^1 \frac{\partial F_g}{\partial a_t}(\alpha a)\, d\alpha ,
  \qquad
  \sum_t e_g(t) = P_g - b_g .
  \label{eq:evidence}
\end{equation}
The sum is the fundamental theorem of calculus, $\sum_t e_g(t) = F_g(a) - F_g(0)$ with $F_g(0) = 0$: the words'
evidence accounts for the answer above its no-information share, with nothing left over. The integral is computed
with no quadrature error by Gauss--Legendre quadrature: $m$ nodes integrate polynomials up to degree $2m-1$ exactly,
and the integrand has degree 199, so 100 nodes suffice. The bound is loose: on ten test pairs the evidence with
30 nodes already agrees with that of 400 nodes to $2 \times 10^{-13}$, so 100 nodes are a guarantee rather than a fine
tuning. We call $e_g(t)$ the \emph{Born evidence per word} (BEW). Two remarks keep the claim in its place. Integrated
gradients of any smooth model is complete in the limit of many nodes, but in general no fixed number of nodes
guarantees exactness. For our softmax baseline's logit over its embeddings, 4,096 nodes bring the median gap from
$0.1$ (256 nodes, seed 0, on the first 200 test pairs; on the 1,000 sampled pairs of Table~\ref{tab:esnli}, Section~\ref{sec:esnli}, the median is 0.02--0.04 over seeds) to
$10^{-7}$ (largest $0.03$) on those 200 pairs and keep 94\,\% of the top three words; for the word-gated probability
of the like-for-like comparison, convergence is checked on all 1,000 pairs and three seeds in
Section~\ref{sec:esnli}. What the polynomial form adds is that a finite rule, 100 nodes here, is exact. And BEW explains the answer with the denominators held at the input, along the integration path;
deleting a word recomputes them, which is why we test the evidence by deletion separately (Section~\ref{sec:esnli}). Where the
Born evidence maps of Seeing Why~\cite{seeingwhy} explain a quadratic form in closed form, one forward and one adjoint
pass, BEW explains a model in which the words act on each other through attention; with the denominators held it stays exact, by quadrature. In double precision the gap is at most $10^{-13}$ on 1,000 test pairs in every seed (median at most $1.4 \times 10^{-14}$). A word's evidence can be negative (it argues against the class). Deleting a word means not sending it: its amplitude is set to zero and the others become $1/\sqrt{n-1}$, as for a
pair of $n-1$ words (Section~\ref{sec:esnli}). Integrated gradients and the Shapley value of the held polynomial coincide for a
quadratic form, as in Seeing Why~\cite{seeingwhy}; at higher degree they share a monomial differently~\cite{lundstrom2023ig}, and we use
integrated gradients. Holding the denominators changes what is explained, not what is computed: the answer is the
model's own, and the held denominators are those of the input. Exact means exact for this function, with every
normaliser held; a camera's estimate of $P_g$ from finitely many photons is not exact, and its noise is a separate
question (Section~\ref{sec:compile}). On light, the row sums are formed from detected readings, the camera's total is detected and $\sum_t a_t$ is set at the modulator; on the device,
the attention row sums are measured and calibrated (Section~\ref{sec:device}), the rest computed. Appendix~\ref{sec:freeden} explains the same model with the denominators free, and by the exact
Shapley value of the deletion game.

Exactness says that each word's evidence is computed without quadrature error, and with it that the evidence adds up
to the answer above its share; it does not say that the evidence is right. For that we need a task in which people
have said which words decide the answer.

\section{Natural language inference with exact evidence}
\label{sec:esnli}

\subsection{The task: does one sentence follow from another?}
\begin{figure}[t]
  \centering
  \includegraphics[width=\linewidth]{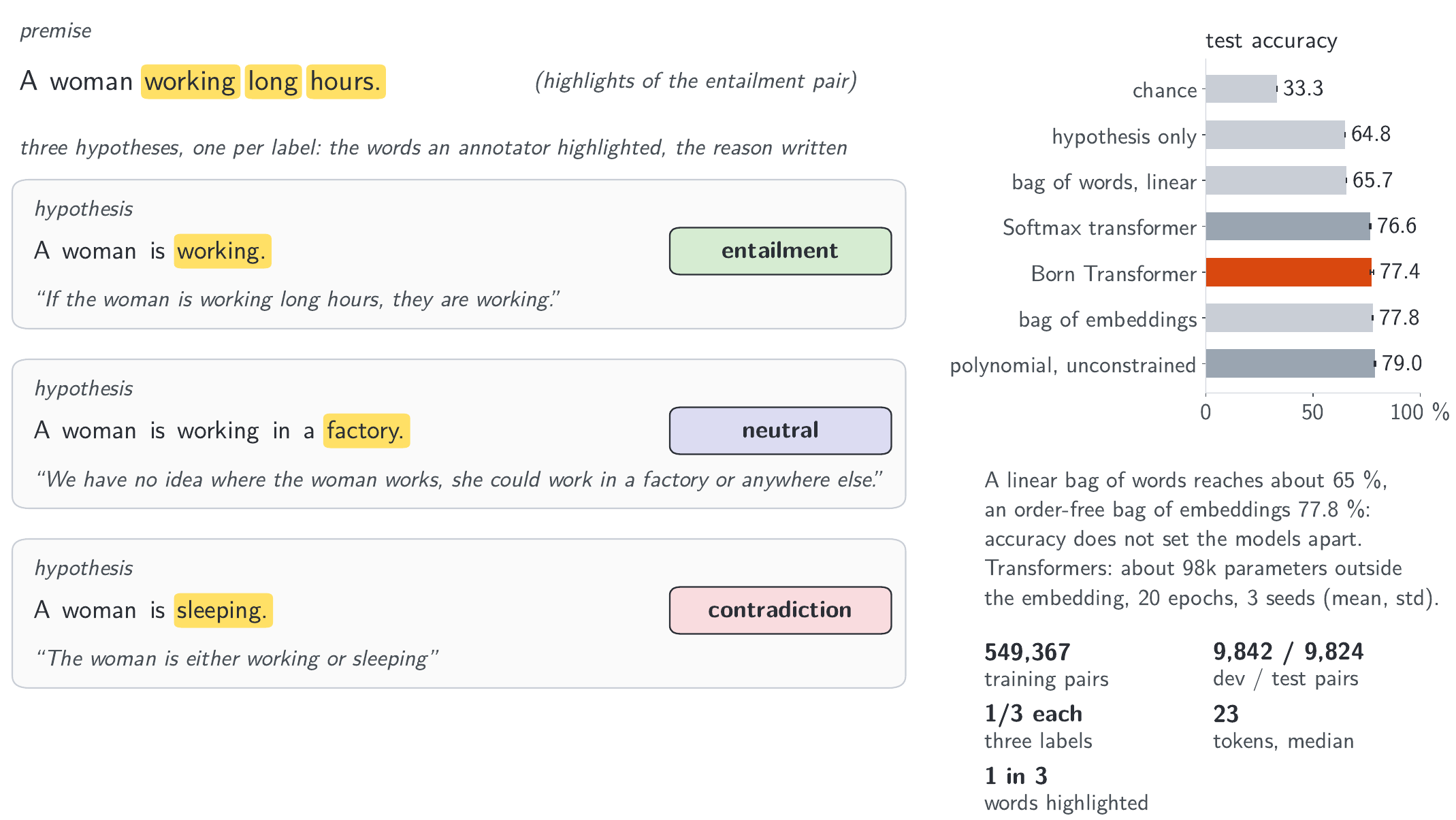}
  \caption{The task: one premise of the e-SNLI test set with its three hypotheses, one per label, the words the first
  annotator highlighted and, beneath each, their written explanation. Right: test accuracy of simple baselines (a
  linear bag of words; an order-free bag of embeddings), of the two transformers of the same size and of the
  unconstrained polynomial model (three seeds). Below: the dataset in numbers.}
  \label{fig:task}
\end{figure}

Natural language inference asks whether a hypothesis follows from a premise. Given ``A woman working long hours'',
the hypothesis ``A woman is working'' follows (\emph{entailment}), ``A woman is sleeping'' cannot be true at the same
time (\emph{contradiction}), and ``A woman is working in a factory'' may or may not be (\emph{neutral};
Fig.~\ref{fig:task}). In principle the answer is in neither sentence alone but in how the two relate, though much of it
can be read from the words (Section~\ref{sec:accuracy}). SNLI~\cite{bowman2015snli} collected such pairs about image captions; e-SNLI~\cite{camburu2018esnli} added, for
every pair, the words an annotator highlighted as the reason and a sentence explaining it. Those highlights are what
makes the data useful here: they are an outside account of which words decide the answer, against which a model's
evidence can be checked.

\subsection{Data and protocol}
e-SNLI has 549,367 training, 9,842 dev and 9,824 test pairs, about a third per label; a test pair has 23 tokens at the
median (premise, separator, hypothesis; vocabulary 22,316 words seen three times or more, 22,319 entries with the special tokens), and annotators highlight
about one word in three. A bag of words with logistic
regression reaches at most 65.7\,\% (one bag per sentence; the hypothesis alone gives 64.8\,\%, the known bias of
SNLI~\cite{gururangan2018artifacts,poliak2018hypothesis}). A bag of embeddings, with no word order and no attention, reaches
$77.79\,\%$ under the same protocol (Section~\ref{sec:accuracy}): at this size accuracy does not show that a model relates the words beyond comparing
the two sentences' means, and what sets the models apart below is their evidence. We compare three models of the same size (98--100k parameters
outside a 1.43M embedding), each with two blocks, two heads and a pooled readout: the Born Transformer; a softmax
transformer (real $d = 64$, softmax attention, GELU~\cite{hendrycks2016gelu} feed-forward of width 256, LayerNorm~\cite{ba2016layernorm}, a linear head); and
Poly-sim-born, a real-valued polynomial transformer with normalised
attention $(q \cdot k)^2/\sum$ and a bilinear feed-forward but no constraint from measurements (a
simulation reference, not a hardware model). Every model learns its embedding from random initialisation on e-SNLI
alone; no pretrained word vectors are used, so that the Born Transformer's complex embeddings, phases included, are
shaped by its own interference. All were trained for 20 epochs on sentences of up to 64 tokens in batches of 64 with Adam~\cite{kingma2015adam}, learning
rate $2 \times 10^{-3}$ (chosen on the dev split, 9,842 pairs, for the softmax model and for the Born model with the weight squared again, Appendix~\ref{app:details}), cosine decay, word dropout 0.1, three
seeds; the softmax model's evidence is integrated gradients of its logit over the embedded vectors (256 nodes),
which is not exact for it.
Evidence is judged three ways, each on the predicted class: the \emph{gap}, how far the word evidence is from summing
to the explained score; \emph{deletion}, deleting the top 1--5 words by evidence and recording the mean drop of the
answer's probability (AOPC~\cite{samek2017aopc}) and how often the answer changes (flip@$k$), against a random order of deletion; and
\emph{agreement}, the share of the evidence's top $k$ words that annotators highlighted, with $k$ the number of
highlighted words in the pair (chance 0.354). Deleting a word sets its amplitude to zero in the Born Transformer (the other words' amplitudes become $1/\sqrt{n-1}$, as
for a pair of $n-1$ words; a rescaling by $\sqrt{23/22}$, the deletion's at the median length, with no word removed,
moves the probabilities by up to 0.048 on 200 pairs, seed 0) and removes
the token from attention and pooling in the others; in every model the deletion is in place, so positions do not
shift. Every token, the separator and punctuation included, is ranked and can be deleted; the ceiling is over the
same tokens. The highlights are the
union over the test split's three annotators, and every metric uses the same 1,000 test pairs.

\subsection{Accuracy and evidence}
\label{sec:accuracy}

\begin{figure}[!htb]
  \centering
  \includegraphics[width=\linewidth]{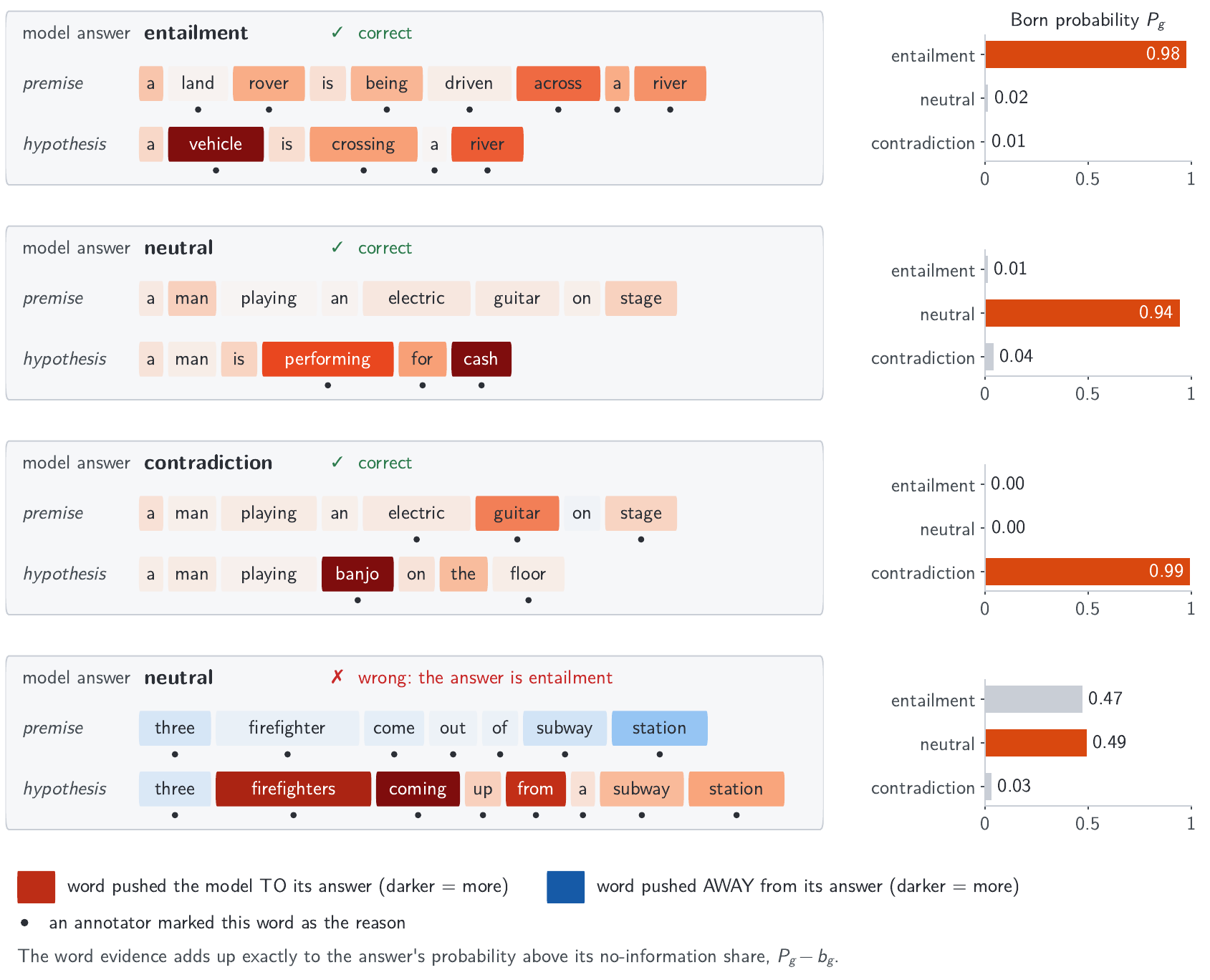}
  \caption{Born evidence per word (BEW) for four short e-SNLI test pairs (at most 18 tokens, two or more words
  highlighted): the first right answer of each class in test order and the first wrong one. Each word is shaded by its evidence for the model's answer (red for, blue against; scaled
  within the pair); dots mark the words annotators highlighted; bars give the Born probabilities.}
  \label{fig:examples}
\end{figure}

Table~\ref{tab:esnli} sets the three models side by side: accuracy, the completeness of their evidence, deletion
and agreement; Fig.~\ref{fig:examples} shows the evidence on four test pairs. The gap is that of the Born
Transformer's probability, of Poly-sim-born's logit (a polynomial it explains exactly) and of the softmax model's
logit, which it explains by integrated gradients with 256 nodes, not exact for it (its largest gap, 47, is on a pair
where the explained difference is near zero). Deletion is measured on $P_g$ itself for the Born models and on the
softmax probability for the others; versions 1--2 of this paper measured the Born models' AOPC on Seeing Why's
headroom softmax of $P_g$, which raised it by about two fifths (0.618 for the fourth-power model, random 0.125), the
flips only in the third digit. The bag of embeddings averages 64-dimensional embeddings over each sentence, $u$ and
$v$, and feeds $[u, v, u \odot v, |u-v|]$ to one hidden layer of 300 units (78k parameters besides the embeddings;
same protocol; \texttt{bag\_baseline.py}).

\begin{table}[h]
\centering\footnotesize
\setlength{\tabcolsep}{2.8pt}
\begin{tabular}{lc@{\hspace{6.5pt}}cccccc}
\toprule
 & test & dev & \multicolumn{2}{c}{gap} & AOPC & flip@1 & agreement \\
\cmidrule(lr){4-5}
model & (\%) & (\%) & median & largest & (random) & (random) & (annot.) \\
\midrule
\rowcolor{ours} Born Transformer & $77.45\pm0.91$ & 77.55 & $1.4{\times}10^{-14}$ & $9.8{\times}10^{-14}$ & 0.446 (0.073) & 0.44 (0.06) & $0.520\pm0.003$ \\
\quad attention $|\braket{q}{k}|^4$ & $76.98\pm0.17$ & 77.30 & $6{\times}10^{-13}$ & $8{\times}10^{-12}$ & 0.445 (0.066) & 0.45 (0.07) & $0.510\pm0.005$ \\
Softmax & $76.62\pm0.36$ & 76.36 & $0.04$ & 47 & 0.321 (0.059) & 0.31 (0.05) & $0.468\pm0.003$ \\
Poly-sim-born & $78.98\pm0.31$ & 78.89 & $1{\times}10^{-14}$ & $5{\times}10^{-12}$ & 0.501 (0.065) & 0.45 (0.06) & $0.527\pm0.011$ \\
Bag of embeddings & $77.79\pm0.26$ & 77.71 & -- & -- & -- & -- & -- \\
\bottomrule
\end{tabular}
\caption{e-SNLI, 20 epochs, three seeds: accuracy and agreement as mean $\pm$ sample standard deviation, the others
as means. Gap: how far the word evidence is from summing to the explained score, relative to $|F(a) - F(0)|$ (worst
seed's median / largest over the 1,000 pairs). AOPC and flip@1 on the predicted class, a random order of deletion in
brackets; agreement: share of the top $k$ words that annotators highlighted (chance 0.354). Second row: the attention
weight squared again, $|\braket{q}{k}|^4$ over its row sum (degree 648 against 200).}
\label{tab:esnli}
\end{table}

The Born Transformer is level with the softmax model (Table~\ref{tab:esnli}; four test pairs and their evidence
in Fig.~\ref{fig:examples}, where the wrong case shows why it is wrong: ``firefighters'' and ``coming'' in the
hypothesis pull towards neutral, and the premise, which should decide for entailment, pushes against it but too
weakly---the answer is nearly a tie, 0.49 against 0.47); the point is not to beat
it but to
show that a model built from quadratic detections reaches accuracy comparable to a softmax model of the same size in
these runs (three seeds, overlapping spreads) while making the evidence exact. Deletion needs care:
the Born model's probabilities move somewhat more under any deletion, random ones included (AOPC 0.073 against
0.059), so its AOPC is larger for that reason too---by evidence it is 0.37 above random against 0.26, and 6.1 times
random against 5.5. Whether the answer changes does not depend on the probability scale: deleting the word the evidence ranks first changes it in 44\,\% of pairs against 31\,\% (the softmax model's integrated gradients of its logit; 32\,\% for its gated probability, below), while a random word changes it in 5--6\,\% for both. This top-word gain is the model's, not BEW's: the Born
Transformer's own gradient $\times$ input, one backward pass with nothing held, changes the answer by its first word in
43\,\% of pairs (0.432 against BEW's 0.435 over seeds and pairs, Table~\ref{tab:attn}), while the softmax model's
gives 14\,\%. Integrated gradients with every denominator recomputed along the path reach only 33\,\%, because that
path starts at a meaningless limit (Appendix~\ref{sec:freeden}), not because holding raises the top word. Where holding does help is the other end and the agreement:
words the evidence ranks last should not matter. Deleting the five lowest changes the Born model's answer in 1\,\%
of pairs (its probability rises, AOPC $-0.08$: the words ranked last argue against the answer), against 10\,\% for
the softmax model. The Born model's top words also agree more
with the annotators than the softmax model's evidence does ($0.520\pm0.003$ against $0.468\pm0.003$ over seeds;
0.57 against 0.50 on content words, Table~\ref{tab:attn}),
though the softmax model's attention-based rankings agree more still (see below). Poly-sim-born, free of the measurement
constraints but with exact evidence of its own, is one and a half points more accurate. Squaring the attention weight
again (second row of Table~\ref{tab:esnli}) changes neither the accuracy nor the deletion scores beyond the spread of the
seeds, at 3.2 times the degree.

\paragraph{Like for like, and attention.} The comparison in Table~\ref{tab:esnli} gives each model its usual
evidence: for the softmax model, integrated gradients of its logit over the embedded vectors. To compare like for like,
Table~\ref{tab:attn} adds, on the same 1,000 pairs, rankings that every model can have: integrated gradients of the
softmax model's probability over one gate per word---the path of BEW---the attention each
word receives (averaged over blocks, heads and the words that look at it), attention $\times$ gradient (the gradient-weighted attention map $\mathrm{E}_h[(A \odot \partial P_g/\partial A)^+]$
of Chefer et al.~\cite{chefer2021generic}, with the gradient of the probability in place of the class output, summed over blocks and over the words that look at each word, without
their propagation across layers), gradient $\times$ input~\cite{shrikumar2016blackbox}, and leave-one-out deletion~\cite{li2016erasure}, which comes within 0.3 points of the ceiling: the share of pairs in which some single
deletion changes the answer. Table~\ref{tab:attn} gives every score; the text keeps to the comparisons. Its intervals
are 95\,\% bootstraps over pairs; resampling the seeds as well widens them to about $\pm0.03$ (up to $\pm0.05$) for
flip@1 and $\pm0.015$ (up to $\pm0.025$) for agreement, and of the comparisons in the text only Born evidence against
Born attention on agreement, $+0.01$ ($-0.01$ to $+0.03$), is not resolved. On correctly answered pairs only, every
agreement but one rises, by up to 0.03, and the comparisons hold. The content-word column takes punctuation and the
separator out of both ranking and highlights, since the highlights mark punctuation attached to a highlighted word
(chance 0.357, intervals about $\pm0.012$). Gradient $\times$ input is the model's own, its denominators free; the
softmax model's gated integrated gradients use 4,096 nodes (median gap $5\times10^{-8}$ in the worst seed, largest
0.045; the content column 512); attention rollout~\cite{abnar2020rollout} gives agreement 0.52, 0.55 and 0.42 (Born
Transformer, softmax, Poly-sim-born).

\begin{table}[!htbp]
\centering\footnotesize
\setlength{\tabcolsep}{2.7pt}
\begin{tabular}{llccccc}
\toprule
model & ranking & flip@1 & of & flip@5, & agreement & content \\
 & & & ceiling & lowest & & \\
\midrule
\rowcolor{ours} Born Transformer & evidence (BEW) & 0.44 [0.41, 0.46] & 82\,\% & 0.012 & 0.52 [0.51, 0.53] & 0.57 \\
 & attention & 0.31 & 58\,\% & 0.087 & 0.51 & 0.57 \\
 & attention$\times$gradient & 0.38 & 71\,\% & 0.074 & 0.54 & 0.61 \\
 & gradient$\times$input & 0.43 & 81\,\% & 0.055 & 0.47 & 0.50 \\
 & leave-one-out & 0.53 [0.50, 0.55] & 99\,\% & 0.005 & 0.49 & 0.53 \\
\midrule
Softmax & IG, logit, embeddings & 0.31 & 56\,\% & 0.103 & 0.47 & 0.50 \\
 & IG, $P_g$, word gates & 0.32 [0.30, 0.34] & 58\,\% & 0.099 & 0.46 [0.45, 0.47] & 0.49 \\
 & attention & 0.34 & 61\,\% & 0.052 & 0.54 & 0.59 \\
 & attention$\times$gradient & 0.40 [0.38, 0.43] & 72\,\% & 0.036 & 0.55 & 0.61 \\
 & gradient$\times$input & 0.14 & 26\,\% & 0.289 & 0.40 & 0.40 \\
 & leave-one-out & 0.55 [0.53, 0.58] & 100\,\% & 0.003 & 0.50 & 0.53 \\
\midrule
Poly-sim-born & evidence (exact) & 0.45 & 83\,\% & 0.017 & 0.53 & 0.57 \\
 & attention & 0.24 & 45\,\% & 0.103 & 0.44 & 0.47 \\
 & attention$\times$gradient & 0.33 & 60\,\% & 0.027 & 0.49 & 0.55 \\
 & gradient$\times$input & 0.43 & 79\,\% & 0.030 & 0.48 & 0.52 \\
 & leave-one-out & 0.54 & 100\,\% & 0.004 & 0.51 & 0.55 \\
\midrule
Bag of embedd. & gradient$\times$input & 0.43 & 70\,\% & 0.058 & 0.47 & -- \\
\quad (seed 0) & IG, $P_g$, embedd. & 0.39 & 64\,\% & 0.073 & 0.48 & -- \\
 & leave-one-out & 0.61 & 100\,\% & 0.009 & 0.48 & -- \\
 & random & 0.09 & 14\,\% & 0.253 & 0.35 & -- \\
\midrule
no model & last token first & 0.06--0.08 & -- & -- & 0.59 & -- \\
 & rarest hyp.\ word first & 0.21--0.24 & -- & -- & 0.58 & -- \\
 & hyp.\ tokens, random & 0.09--0.10 & -- & -- & 0.50 & -- \\
any & random & 0.05--0.06 & 10--11\,\% & 0.19--0.20 & 0.35 & 0.36 \\
\bottomrule
\end{tabular}
\caption{Every ranking on the same 1,000 test pairs, pooled over three seeds (the bag of embeddings: seed 0), with
95\,\% intervals over pairs in brackets (about $\pm0.02$ for flip@1 and $\pm0.01$ for agreement where not shown).
flip@1: deleting the top word changes the answer; of ceiling: flip@1 over the share of pairs that some single deletion
changes; flip@5, lowest: deleting the five ranked last changes it; agreement: share of the top $k$ that annotators
highlighted; content: the same on content words. No model: flip@1 with the five runs' own answers.}
\label{tab:attn}
\end{table}

The three transformers' ceilings are alike, so none of them is simply easier to flip. Explained by integrated gradients of its probability along the same path, the softmax model's evidence
falls 12 points below the Born Transformer's on flip@1 (95\,\% interval 6--17 when seeds and pairs are both
resampled); its best ranking short of trying every deletion, attention $\times$ gradient, is 3.5 points below
(0.1--7), a difference that rests on three seeds (BEW's seed 0, the model of the device runs, is level with it at
0.408). The ranking comparison is robust to increasing the quadrature resolution: with 4,096 nodes the softmax model's integrated gradients of its gated probability have a median gap of $10^{-10}$ over seeds and pairs ($5\times10^{-8}$ in the worst seed; the largest 0.045), 99.7\,\% of the per-pair outcomes
are those of 512 nodes, and on the pairs whose gap is below $10^{-2}$, $10^{-3}$ or $10^{-4}$ in every seed (991, 944
and 815 pairs) its flip@1 stays at 0.30--0.32. These checks indicate that integration error does not explain the
observed ranking gap. Nor does its baseline: with every gate at zero the softmax model gives one fixed answer, neutral in every seed
(probability 0.78--0.98), so its gated evidence sums to $P_g$ above that answer rather than above the share $b_g$; yet its
flip@1 is about the same on the pairs it answers neutral and on the others (0.33 and 0.31, mean of the seeds). Gradient $\times$ input of the model
itself ($a_t\,\partial P_g/\partial a_t$, the denominators free) ranks the first word fairly
well in the polynomial models but not the last ones, and agrees less with the annotators; in the softmax model it
fails, its five lowest changing the answer more often than five random words. With the denominators held, as in BEW,
it ranks the words much as BEW does (rank correlation 0.94--0.97, the same top word in 83--90\,\% of pairs; flip@1
0.41 against 0.44, the five lowest 1.3 against 1.2\,\%, means of the seeds): $F_g(\alpha a)$ stays near zero along most
of the path and rises near its end, roughly as $\alpha^k$ with $k$ about 30--45 (the median over pairs of
$\sum_t a_t\,\partial F_g/\partial a_t$ over $\sum_t e_g(t)$), for which integrated gradients are gradient $\times$ input
divided by $k$. The difference from the free gradient $\times$ input in Table~\ref{tab:attn} is at the low end and in agreement
(five lowest 1.2 against 5.5\,\%, agreement 0.52 against 0.47), not at the top (0.44 against 0.43); holding gives
that, and integrating along the path adds mainly the exact sum, not the ranking. The softmax model's
denominators cannot be held in the same way: with its attention row sums, LayerNorm scales and output sum held at
the input, its score along the path overflows double precision on most pairs (exponentials of exponentials through
the two blocks), and its integrated gradients are finite and add up (to $10^{-3}$, 1,024 nodes) on 4.5\,\% of the pairs
only (42, 39 and 55 in the three seeds). On those pairs they rank worse than its gated integrated gradients with the
denominators free (flip@1 0.24 against 0.33; the five lowest change the answer in 12 against 9\,\%), and BEW reaches
0.41 and 2.4\,\% on the same pairs. Holding the denominators helps the low end of the ranking where the held model is a polynomial;
on the subset where the softmax model's held integrated gradients can be computed, it did not improve its ranking.

In the softmax model attention looks most like the annotators' reason: by attention $\times$ gradient its top words agree with them best of all (in the Born Transformer attention and evidence agree with them about equally, in Poly-sim-born attention less).
On content words alone (last column) the agreements rise by up to 0.06 and keep these orderings. It is a weaker account of the answer: deleting the word attention ranks first
changes the answer far less often than deleting the evidence's first word, in the Born Transformer (31 against
44\,\% of pairs) and in Poly-sim-born (24 against 45\,\%). Looking like a
person's reason and accounting for the answer are different properties, and only the evidence is checked against the
answer it explains.

Agreement is also weak evidence by itself. Two rankings that need no model agree with the annotators better than
any model's ranking on these 1,000 pairs: the last token first, 0.585, and the hypothesis's rarest content word
first, 0.582; the hypothesis's tokens in random order, 0.501, are level with the models' evidence
(\texttt{esnli\_baselines.py}). The annotators mostly mark late, rare
words of the hypothesis, and agreement rewards a ranking for preferring them; it does not show that any model's
evidence follows the annotators better than word position does, and the agreements in
Table~\ref{tab:attn} are best read against these baselines, not against chance. Deletion separates the rankings
where agreement does not: the rarest hypothesis content word first changes the answer in 21--24\,\% of pairs and the
last token first in 6--8\,\% (with the three seeds' answers), BEW in 41--47\,\% and the softmax model's
integrated gradients in 31\,\% (three seeds, Table~\ref{tab:esnli}). BEW's first token
is also not always a word: for seed 0, the model of the device runs, it is the separator in 21\,\% of pairs (seeds 1
and 2: 6 and 9\,\%, with punctuation in 8 and 3\,\%), and the separator's signed evidence is 12\,\% of the total absolute evidence (seeds 1 and 2: 8 and 7\,\%). Deleting a
first-ranked separator or punctuation mark changes the answer in 11--18\,\% of those pairs, a first-ranked word in
47--51\,\%; this is part of the gap between BEW and leave-one-out.

An attention-free model does as well at the top. The bag of embeddings of Table~\ref{tab:esnli} (seed 0), explained
by its own gradient $\times$ input, changes the answer by its first word in 42.5\,\% of the same pairs, by integrated
gradients of its probability (1,024 nodes) in 39.0\,\%, and by leave-one-out in 60.6\,\%, its ceiling
(\texttt{bag\_evidence.py}). In absolute flip@1 it is level with BEW; BEW is ahead relative to the ceiling (82
against 70\,\%) and at the low end (its five lowest change the answer in 1.2 against 5.8\,\%), and the bag's ceiling
is higher (0.61 against 0.53), so it is easier to flip. What BEW offers that these rankings do not is its exact sum
and the low end of its ranking, not a better first word. Its ranking as a whole is not that of deletion: BEW against its own
leave-one-out has a median rank correlation of 0.45--0.51 over seeds (Poly-sim-born 0.67), so ``accounts for the
answer'' is a statement about the first word and the sum.

\paragraph{Which part gives the evidence its quality.} Poly-sim-born, the polynomial model without the measurement
constraints, lets one part be swapped at a time (Table~\ref{tab:ablation}). With a GELU feed-forward in place of the
bilinear one, the attention still a ratio of squared overlaps, the accuracy holds (77.1\,\%), but the model is no
longer a polynomial, its integrated gradients no longer sum to the score (median gap $4\times10^{-3}$, largest 6), and
the evidence weakens on every count: the first word changes the answer in 27\,\% of pairs instead of 45\,\%, the five
lowest in 11\,\% instead of 2\,\%, and agreement with the annotators falls from 0.53 to 0.46. With softmax attention
the model loses about 13 points of accuracy (78.98 to 65.65\,\%)---without LayerNorm, which the softmax transformer of Table~\ref{tab:esnli}
has, it trains poorly here, at the same learning rate, not tuned for it, so this is not a verdict on softmax attention---and its evidence ranks words worse than
chance at the low end. As that variant trains poorly, the GELU variant is the comparison that bears weight: in it the quality of the
evidence goes, in this ablation, with the whole model being one polynomial (its attention row sums are held in both
variants), not with one part of it.

\begin{table}[h]
\centering\footnotesize
\setlength{\tabcolsep}{3pt}
\begin{tabular}{llccccc}
\toprule
attention & feed-forward & test (\%) & gap (median) & flip@1 & flip@5, lowest & agreement \\
\midrule
\rowcolor{ours} $(q{\cdot}k)^2$ / row sum & bilinear & $78.98\pm0.31$ & $10^{-14}$ & 0.45 & 0.02 & $0.53\pm0.01$ \\
$(q{\cdot}k)^2$ / row sum & GELU & $77.13\pm1.42$ & $4\times10^{-3}$ & 0.27 & 0.11 & $0.46\pm0.02$ \\
softmax & bilinear & $65.65\pm0.44$ & not complete & 0.15 & 0.25 & $0.37\pm0.01$ \\
softmax & GELU & $65.47\pm0.25$ & not complete & 0.17 & 0.23 & $0.39\pm0.01$ \\
\bottomrule
\end{tabular}
\caption{Poly-sim-born (first row) with its attention, its feed-forward or both replaced; same size, training and
test pairs as Table~\ref{tab:esnli}, three seeds. Evidence: exact quadrature for the polynomial model, integrated
gradients with 256 nodes otherwise. flip@1: deleting the word ranked first changes the answer (random 0.05--0.06);
flip@5, lowest: deleting the five ranked last does (random 0.17--0.20). Agreement: chance 0.354.}
\label{tab:ablation}
\end{table}

The model was trained on a computer, and nothing in it refers to a device. What does it take to run it on one?

\section{One set of weights, two realisations}
\label{sec:compile}
Compiling it is a matter of where the scale goes.

\begin{figure}[!htbp]
  \centering
  \includegraphics[width=\linewidth]{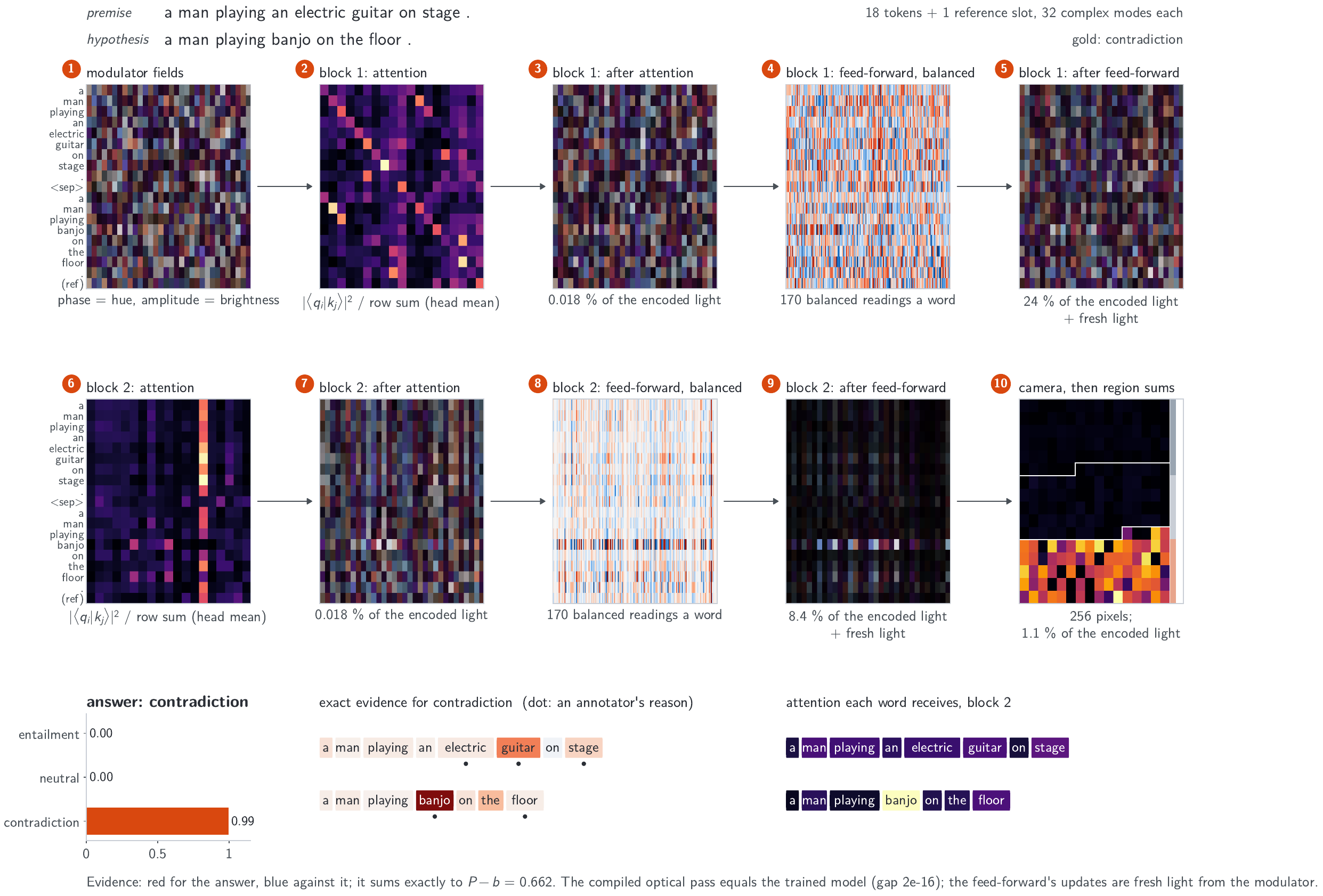}
  \caption{One test pair through the compiled optics, stage by stage: the modulator's fields (phase as hue), each
  block's attention and field after attention, the balanced readings of the measured feed-forward and the field after
  it (with the share left of the light the first modulator encoded for this pair; the $10^{-3.2}$ of the text is the
  share of the source light, the mean over 512 pairs), and the camera with its three class regions; below, the
  answer, the exact evidence of each word and the attention each word receives in the last block.}
  \label{fig:optical}
\end{figure}

\paragraph{Optics.} Every fixed matrix is divided by its largest singular value, so that it is a
contraction---a block of a unitary on twice the modes, built from Mach--Zehnder interferometers and attenuators~\cite{miller2013selfconfiguring}. Every word field is
scaled to norm at most one. Splitting a state into its query, key, value and residual copies, fanning a key out to
the slots it meets, and combining the residual are passive beam splitters, each losing a known share of the light.
None of this changes the answer: the attention weights~\eqref{eq:attention} and the readout~\eqref{eq:readout} are
ratios and do not see a common scale, and the only place where scale matters, the measured
feed-forward~\eqref{eq:mlp}, is followed by a classical gain on measured numbers that puts the lost scale back before
the modulator writes $W_3 r$. The build is optoelectronic, not passive throughout: between its interferometers
are detectors, a classical gain and fresh modulators. The attention weights are written the same way: the controller squares and divides
the detected overlaps and reprograms attenuators for each input, so each block takes at least two optical shots---one to read the
overlaps, one to send the value fields through the written weights---or the value fields wait in an optical delay;
whether the slots are laid out in space or in time we leave open. The photon numbers below count one pass with the value
fields held; re-sending them multiplies the light by the number of shots, and the feed-forward write-backs add fresh light.
The parts that are sensitive to phase are the coherent value sums $\sum_j w_{ij} v_j$, the residual and slot combiners
and the two arms of each balanced reading; $|\braket{q_i}{k_j}|^2$ does not see a phase common to $q_i$ and $k_j$, only
the $90^\circ$ between its two quadrature readings. The compiled pass equals the trained model to $2 \times 10^{-15}$ (the largest difference on 512 test pairs); what compiling
costs is light, not accuracy (Fig.~\ref{fig:optical}). With shot noise at every detector, the first 512 test pairs (80.1\,\% noiseless) keep 80.3\,\% on average over 20 draws of the noise (lowest 79.5, highest 81.1\,\%) at
$10^{10}$ source photons per sentence (shared among all its detectors), 80.2\,\% (79.9--80.5\,\%) at $10^{12}$ and
68.5\,\% (65.6--70.9\,\%) at $10^{8}$; with finite precision instead, they keep 80.1\,\% at 10 bits per quadrature of every encoded field and per attention weight and
fixed-matrix entry (the converters that read the measured numbers are not modelled); $10^{-3.2}$ of the source light reaches the camera.

\paragraph{Qubits.} The same weights run on qubits without copies of a state: every reading prepares its states
afresh. Each overlap $\braket{q_i}{k_j}$ is a Hadamard test~\cite{cleve1998revisited,aharonov2006jones} on five qubits---the ancilla and the four qubits that hold the 16 modes of one head, with
$q_i$ and $k_j$ normalised:
\begin{equation}
  \frac{\ket{0}\ket{q_i} + \ket{1}\ket{k_j}}{\sqrt2}
  \;\xrightarrow{\;H \text{ on the ancilla}\;}\;
  P_0 - P_1 = \mathrm{Re}\braket{q_i}{k_j},
  \label{eq:hadamard}
\end{equation}
and, with a phase $-i$ on the second branch, $\mathrm{Im}\braket{q_i}{k_j}$; the norms are put back classically.
The attention needs only the Born weight $|\braket{k_j}{q_i}|^2$, and one circuit on four qubits returns it: prepare
$\ket{q_i}$, undo the preparation of $\ket{k_j}$ (both normalised; the norms are put back classically, as for the
Hadamard test) and measure all four,
\begin{equation}
  U_{k_j}^\dagger U_{q_i}\ket{0000} \;\longrightarrow\; P(0000) = |\braket{k_j}{q_i}|^2 ,
  \label{eq:cu}
\end{equation}
a compute--uncompute circuit, with no ancilla and one circuit per overlap instead of two. Here
the state $(\ket{0}\ket{q_i} + \ket{1}\ket{k_j})/\sqrt2$ is prepared directly from the classically known vectors, and
that preparation is most of the circuit's cost. In the same way the feed-forward readings~\eqref{eq:mlp} would be
Hadamard tests, the combiners linear combinations of unitaries~\cite{childs2012lcu} and the readout one more circuit; the costs are shots
instead of light, and the arithmetic is the same. A sentence of $n$ words (with the reference slot, $n+1$ slots)
needs $2 \times 2 \times (n+1)^2 \times 2 = 8(n+1)^2$ overlap circuits for both blocks, heads and the real and
imaginary parts ($4(n+1)^2$ by compute--uncompute), plus calibrations when the attention row sums are wanted for the evidence: for the seven-token pairs
of Section~\ref{sec:device}, 512 overlap circuits and 64 calibration circuits as deep as the overlaps (32 per block).

\paragraph{An outside check of the optical build.} Every fixed matrix of the compiled pass (15 matrices, each counted once: 130k Mach--Zehnder
interferometers, uncalibrated) was built from a full singular value decomposition as $s\,L\,\mathrm{diag}(S)\,R^\dagger$, with $L$ and $R^\dagger$
full square meshes (a $170 \times 32$ matrix gets a 170-mode mesh on one side) in the layout of the Clements decomposition, found by its
reference implementation~\cite{clements2016optimal}. Rebuilt from the mesh phases, the
model gives the same answers on 1,000 test pairs (79.1\,\%, every answer kept, $|\Delta P| \le 10^{-14}$).
With independent phase errors per interferometer (eight draws each), accuracy falls by about 1.4 points at 0.01\,rad (76.9--78.8\,\%,
mean 77.7), to 71.1--76.6\,\% (mean 74.4) at 0.02\,rad and collapses at 0.05\,rad: the 170- and 256-mode
meshes are deep. A thin decomposition of the rectangular matrices would need fewer interferometers and would likely
tolerate larger phase errors; we did not build it. Not counted are the parts that change with every input and grow with
the square of the sentence length---the balanced pairs of the overlaps and the attenuators of the attention
weights---nor the input and write-back modulators.

On light, the measurements inside each block are cheap: a detector reads them and a modulator writes the next field.
On qubits each of them is a separate circuit run many times, but a small one, a Hadamard test on five qubits. We ran
them under a noise model and on a device.

\section{On a quantum computer}
\label{sec:device}
\begin{figure}[tp]
  \centering
  \includegraphics[width=\linewidth]{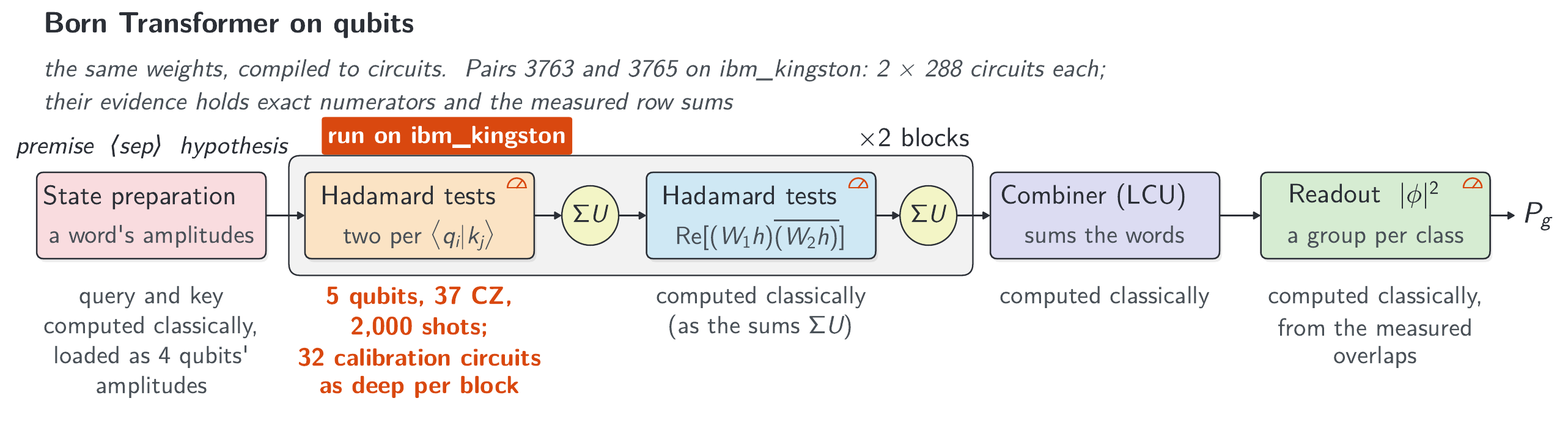}
  \caption{The Born Transformer on qubits, stage by stage, and beside each stage what the Hadamard-test runs of pairs 3763
  and 3765 on ibm\_kingston did with it: the attention read on the device, with calibration circuits as deep, the rest computed
  classically; the evidence is hybrid, exact numerators with the measured row sums.}
  \label{fig:bornqubits}
\end{figure}

Every device run of this section uses the Born Transformer of Section~\ref{sec:esnli} (seed 0, 20 epochs), its
attention read on the device and the rest computed classically (Fig.~\ref{fig:bornqubits}). The coherent attention
layer of Section~\ref{sec:coherent}, a separate model with no measurement before the end, is simulated only.

\paragraph{Under a noise model.} With every overlap of both blocks read by Hadamard tests (10,000 shots) under a
noise model of ibm\_kingston, the Heron r2 device of the runs below (qiskit-aer, FakeKingston; gate and readout
errors, $T_1$, $T_2$), the answer was the same as the noiseless model's on all of the first 50 test pairs
($|\Delta P|$ median 0.006, at most 0.054; 266,336 circuits, median 39 two-qubit gates). In a hybrid form---exact
numerators on the computer, the attention row sums measured and corrected for visibility---the evidence stays
complete, and on 20 pairs it correlates with the exact evidence at a median of 0.997 (lowest 0.988), shares 95\,\%
of its top words, and agrees with the annotators as well (0.54 against 0.54). The noise model is milder than the
device: on the two device pairs its circuits keep a visibility of about 0.93, the device's 0.67--0.81 (below).

\begin{figure}[t]
  \centering
  \includegraphics[width=\linewidth]{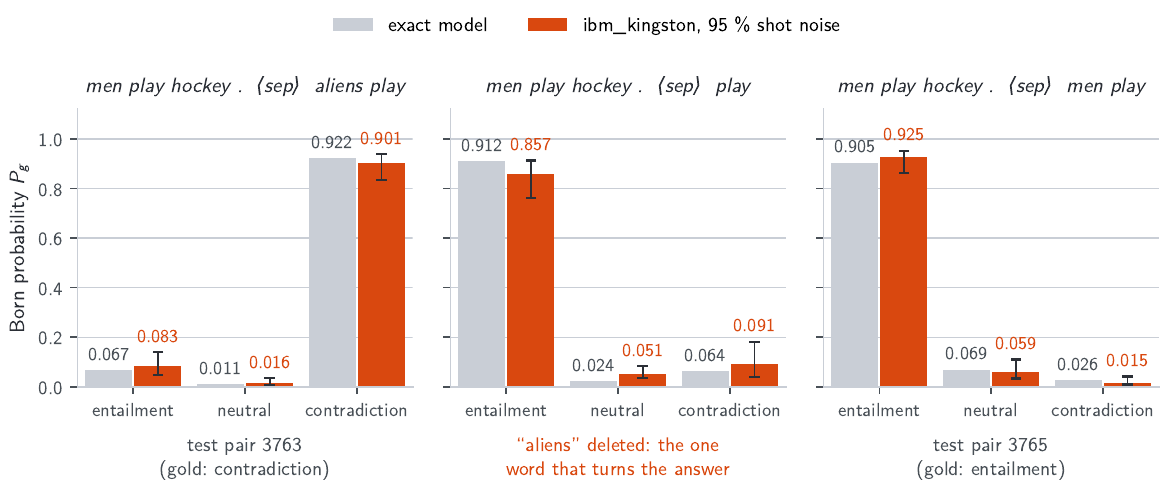}
  \caption{The runs on ibm\_kingston, only the attention overlaps on the device: each run's answer against the exact
  model, with 95\,\% shot-noise intervals. Left: pair 3763; middle: the same pair without ``aliens'', the one word
  whose deletion changes the noiseless model's answer; right: pair 3765.}
  \label{fig:device}
\end{figure}

\paragraph{On ibm\_kingston.} The two shortest test pairs, seven tokens each, ran on this Heron r2 device: pair 3763
(its index in the e-SNLI test split), ``men play hockey . / aliens play'' (contradiction), and pair 3765, ``men play
hockey . / men play'' (entailment). They were chosen by that rule alone, as the pairs that need the fewest circuits, before the first
device run, which was of the earlier fourth-power model (Appendix~\ref{app:details}); the runs of the square model
reported here reuse them. For each pair every $\braket{q_i}{k_j}$ of both blocks was read by a five-qubit Hadamard
test: per block, 256 overlap circuits (the real and imaginary parts of 128 overlaps) and 32 calibration circuits,
Hadamard tests of random state pairs of known overlap that compile to the same 37 CZ gates as the overlaps, in the same
job; $2 \times 288$ circuits per pair, 2,000 shots each, 310 quantum seconds. Only these overlaps ran on the device.
The encoding, the projections, the normalisation of the weights, the feed-forward and the readout were computed
classically between and after the two jobs, and the vectors whose overlaps the device read were known classically:
the run shows that the compiled attention holds up on hardware, not that the device computes something a computer
could not. The measured overlaps followed the ideal ones (correlation 0.992--0.993 over the 256 readings of each
block), scaled by a visibility, the slope fitted to each block's calibrations, of 0.77 and 0.67 for pair 3763 and 0.81
and 0.79 for pair 3765, with a readout offset of at most 0.005, which is taken off every overlap. The visibility
cancels in each attention row (Eq.~\eqref{eq:visibility}), and the device's answer followed the noiseless model's
(Table~\ref{tab:device}, Fig.~\ref{fig:device}). The intervals redraw each circuit's 2,000 shots (1,000 redraws), which measures shot noise within each run, not the
spread between repeated runs (pair 3763's two blocks started three and a half hours apart); in every redraw the device's answer was the same.

\begin{table}[h]
\centering\footnotesize
\setlength{\tabcolsep}{4pt}
\begin{tabular}{llcccc}
\toprule
run & word deleted & answer & device & noiseless & 95\,\% interval \\
\midrule
pair 3763 & -- & contradiction & 0.901 & 0.922 & 0.836--0.938 \\
 & ``aliens'' & entailment & 0.857 & 0.912 & 0.762--0.913 \\
\midrule
pair 3765 & -- & entailment & 0.925 & 0.905 & 0.862--0.953 \\
\midrule
\multicolumn{6}{l}{compute--uncompute} \\
pair 3763 & ``aliens'' & entailment & 0.865 & 0.912 & 0.849--0.884 \\
pair 3765 & -- & entailment & 0.877 & 0.905 & 0.841--0.904 \\
\bottomrule
\end{tabular}
\caption{The runs on ibm\_kingston: the probability of the answer, on the device and in the noiseless model, with the
95\,\% shot-noise interval of the device's value; in every redraw the device's answer was the same. Above, Hadamard
tests; below, compute--uncompute circuits. Pair 3763 is
``men play hockey . / aliens play'', pair 3765 ``men play hockey . / men play''. Jobs, circuits and quantum seconds:
Table~\ref{tab:jobs}.}
\label{tab:device}
\end{table}

\begin{table}[!htbp]
\centering\footnotesize
\setlength{\tabcolsep}{3.5pt}
\begin{tabular}{lcccc}
\toprule
 & \multicolumn{3}{c}{answer kept} & evidence corr. \\
\cmidrule(lr){2-4}\cmidrule(lr){5-5}
what the device returns & 3763 & 3765 & deletion & 3763 / 3765 \\
\midrule
\rowcolor{ours} the device (Hadamard tests) & yes & yes & yes & 0.962 / 0.997 \\
uniform attention, both blocks & yes (0.89) & no (C 0.83) & no (C 0.59) & -- \\
random overlaps, both blocks & 72\,\% & 28\,\% & 42\,\% & -- \\
uniform attention, block 2 only & yes (0.69) & yes (0.81) & yes (0.79) & -- \\
random overlaps, block 2 only & 97\,\% & 97\,\% & 99\,\% & -- \\
block 2 overlaps of one size (norms only)$^{a}$ & -- & -- & -- & 0.832 / 0.995 \\
block 2 Haar-random overlaps (median) & -- & -- & -- & 0.821 / 0.983 \\
no visibility correction & -- & -- & -- & 0.62 / 0.80 \\
random row errors, device's size (median) & -- & -- & -- & 0.980 / 0.999 \\
shot noise alone (median) & -- & -- & -- & 0.992 / 0.999 \\
seed 1's own evidence & -- & -- & -- & $0.01$ / $-0.14$ \\
seed 2's own evidence & -- & -- & -- & $-0.43$ / $-0.33$ \\
\bottomrule
\end{tabular}
\caption{What each check can tell (deletion: pair 3763 without ``aliens''). Answer: whether the compiled model (seed 0),
its attention replaced as the row says, keeps the model's answer, or the share of random draws that keep it; in
brackets the probability of that answer, or of contradiction (C). Evidence: correlation with the exact evidence with
block 2, or all row sums, replaced as the row says; the last two rows are other trained models' own evidence.
$^{a}$Every normalised block-2 overlap set to one value, the norms put back classically: no device information, and
not uniform attention, since the weights then follow the key norms.}
\label{tab:checks}
\end{table}

\paragraph{What the answer check can show.} An answer check tells the device from uninformative attention only where
uninformative attention would answer otherwise. Pair 3763's answer says nothing about the device's attention: uniform attention gives it too. On the
other two runs uniform attention answers otherwise, but a device returning random overlaps (independent Gaussian
real and imaginary parts on the present slots) would pass them about a quarter and two fifths of the time (22--29
and 38--44\,\% of the draws over our scripts), and both together, if independent, about one time in ten. The checks
also test block 1 almost alone: with block 1 exact, uniform or random attention in block 2 keeps the model's answer,
so no answer check tests the block-2 jobs, half of the circuits; only the evidence does, and on pair 3763 only (on
pair 3765 block-2 overlaps all of one size, the norms put back classically and no device information,
still give 0.995, though Haar-random
block-2 overlaps fall below the device in every draw on both pairs). The answer checks are
weak evidence on their own; the evidence check below asks more. Table~\ref{tab:checks} gives, for each kind of
uninformative device, whether the compiled model still gives the model's answer and how closely the evidence still
follows the exact one (100 random draws, block 2 alone 200; Haar 100; random row errors 100; shot noise 60; \texttt{device\_power.py}, \texttt{device\_blocks.py},
\texttt{device\_block2\_nulls.py}, \texttt{device\_floor.py}).

On pair 3763 one word decides the answer in the noiseless model: without ``aliens'' it turns to entailment (0.912),
while deleting any other word keeps contradiction (at least 0.748). The device was asked about that deletion
($2 \times 210$ circuits, 228 quantum seconds) and turned to entailment as well. This checks that the circuits
reproduce the model under a deletion; it does not validate the evidence. Its 14 calibrations per block were shallow
self-overlaps (Table~\ref{tab:jobs}), and with them its evidence fails (correlation 0.693, top word wrong): the
square model's own example of why the calibrations must be as deep as the overlaps. With each block's slope taken
instead from its overlaps against their known ideal values (0.803 and 0.794), the evidence follows the exact one
(correlation 0.999, $\ell_1$ error 2.9\,\%; \texttt{device\_rows.py}), but that slope uses the ideal values the
device is meant to measure, so we count it as a check, not as a third result. Every model we tested in simulation changes
its answer without ``aliens'' (the three seeds of the Born Transformer, to entailment in two and to neutral in one;
the softmax transformer's three seeds, Poly-sim-born and the coherent layer, to entailment). And on this pair the evidence does
not rank ``aliens'' first: it ranks the separator first (0.180) and ``aliens'' fourth, level with the full stop (both 0.055), and deleting the
separator keeps contradiction. A likely reading is that the word acts through the others: deleting it removes that interaction whole, while
integrated gradients share it, along the path, among the words that take part. Pair 3763 is one of the pairs, 18\,\%
of those some single deletion changes (Table~\ref{tab:attn}), in which the evidence's first word does not change the
answer.

\paragraph{The evidence on the device.} The evidence asks more of the device than the answer does: the attention row
sums $D_i$ it holds feel a loss of visibility squared (Eq.~\eqref{eq:visibility}), so each measured overlap is divided
by its block's calibrated visibility, and each row sum thereby by its square. With the measured row sums and exact numerators, the evidence of pair 3765 follows the exact
one (Table~\ref{tab:device_ev}, Fig.~\ref{fig:overview}b): correlation 0.997 (95\,\% interval 0.988--0.999 when every
circuit's 2,000 shots are redrawn, calibrations included; 500 redraws), every sign right, the same top word (the
separator) and the same answer in every redraw; for one word, the hypothesis's ``play'', the interval misses the exact
value (0.049 against 0.031--0.047). On pair 3763 it follows less closely: correlation 0.962 (0.880--0.991), every sign
right in 99\,\% of redraws but the same top word in only 57\,\%, since the separator (0.180) and ``hockey'' (0.155)
are close; for ``aliens'' and the separator the interval misses the exact value (0.055 against 0.059--0.128, and 0.180
against 0.155--0.177). That is not shot noise alone, and not the block's calibration either. The slope that a
block's 256 overlaps show against their known ideal values agrees with the calibrated one (0.775 and 0.661 against
0.767 and 0.674; with it the correlation is 0.968), and the row means of the readings scatter by 0.008--0.009, what
averaging sixteen circuits gives. The error per circuit is 0.035--0.039 against shot noise of 0.028--0.032, like the
calibrations' scatter of 1.15--1.33 times the shot noise at their values. Simulated on this pair with its slopes,
shot noise alone gives a median correlation of 0.992 (5th percentile 0.966; pair 3765, 0.999), and adding a Gaussian
of 0.022 per circuit (in the readings' units) gives 0.986 (5th percentile 0.945), which takes in the device's 0.962
(\texttt{device\_errors.py}, \texttt{device\_floor.py}). Independent noise that matches the error per circuit
(0.015 in the readings' units) gives a spread of the row sums of 0.069 (log standard deviation; 5th--95th percentile
0.055--0.087 over 200 draws) against the device's 0.085, and 0.022 gives 0.081 (0.065--0.102) with too large an error
per circuit (0.040--0.045; \texttt{device\_rowsd.py}): one device run cannot tell whether part of the error is shared
within rows. On pair 3765 noise of 0.015--0.022 brackets both (row spread 0.055--0.066 against 0.064). A row fitted on its own
readings, an oracle that absorbs this noise, would reach 0.998. More shots per circuit would narrow the error in
part; this pair feels it more because its second block lost more of its visibility (0.67, against 0.77 in its first
block and 0.79--0.81 on pair 3765). Squaring a shot-noisy overlap adds its variance; taking that bias off the row sums
changes the correlations by about 0.003 at most (0.965 and 0.997). The hybrid evidence is complete for the probability
it explains, $\tilde P_g$: exact numerators over the measured, calibrated row sums. It sums, to $2 \times 10^{-14}$,
to $\tilde P_g - b_g$, with $\tilde P_g = 0.946$ on pair 3763 and 0.903 on 3765, against the device's answers 0.901
and 0.925, which use the measured overlaps throughout, and the exact 0.922 and 0.905. The numerators are exact and
computed classically; the device gives only the row sums the evidence holds, sixteen numbers per block.
\begin{table}[h]
\centering\footnotesize
\setlength{\tabcolsep}{4pt}
\begin{tabular}{lccccccc}
\toprule
pair 3765 & men & play & hockey & . & $\langle$sep$\rangle$ & men & play \\
\midrule
exact BEW & 0.048 & 0.070 & 0.110 & 0.064 & 0.200 & 0.027 & 0.049 \\
ibm\_kingston & 0.043 & 0.067 & 0.117 & 0.064 & 0.208 & 0.029 & 0.038 \\
95\,\%, low & 0.036 & 0.059 & 0.103 & 0.056 & 0.187 & 0.023 & 0.031 \\
95\,\%, high & 0.052 & 0.073 & 0.132 & 0.073 & 0.223 & 0.034 & 0.047 \\
\midrule
pair 3763 & men & play & hockey & . & $\langle$sep$\rangle$ & aliens & play \\
\midrule
exact BEW & 0.024 & 0.091 & 0.155 & 0.055 & 0.180 & 0.055 & 0.029 \\
ibm\_kingston & 0.016 & 0.088 & 0.166 & 0.056 & 0.169 & 0.094 & 0.026 \\
95\,\%, low & 0.002 & 0.077 & 0.140 & 0.050 & 0.155 & 0.059 & 0.019 \\
95\,\%, high & 0.032 & 0.098 & 0.194 & 0.063 & 0.177 & 0.128 & 0.032 \\
\bottomrule
\end{tabular}
\caption{Each word's evidence for the answer (entailment for pair 3765, contradiction for 3763), exact and with the
attention row sums measured on ibm\_kingston (calibration circuits as deep as the overlaps, in the same job); the
95\,\% interval redraws every circuit's 2,000 shots, calibrations included (500 redraws). Every exact value is
positive: every word supports the answer.}
\label{tab:device_ev}
\end{table}

The evidence check is a consistency check, not a test of the device. It does detect a wrong calibration or a wrong
model (Table~\ref{tab:checks}): without the visibility correction the correlations fall to 0.62 and 0.80 (mostly
block 1: left uncorrected in block 1 alone they give 0.911 and 0.796, in block 2 alone 0.935 and 0.991), and the
evidence of the two independently trained seeds correlates with seed 0's between $-0.43$ and 0.01 on these pairs. These two pairs are outliers: over the 760 test pairs on which the three seeds agree, BEW's seed-to-seed
median correlation is 0.79--0.82 (the same top word in 58--60\,\%; the softmax model's integrated gradients, 0.48--0.51
and 35--39\,\%; \texttt{esnli\_seed\_stability.py}), so the evidence read on the device, with the separator first on
both pairs, is seed 0's rather than the sentence's. But unbiased random
errors of each row sum as large as the device's (log standard deviation 0.085 and 0.064 over rows) give median
correlations of 0.980 and 0.999, below the device's in 26 and 29\,\% of 100 draws, and shot noise alone gives 0.9994
(median over 100 simulated pairs, below; on these two short pairs 0.992 and 0.999). What it shows is that, once the visibility is calibrated, the
measured row sums are right to 6--9\,\% per row against the exact row sums of the vectors the device read; their
errors cost the evidence somewhat more than random errors of that size would. In block 2 those vectors come from the
device's own block-1 pass, and their exact row sums differ from the model's by 9 and 14\,\% (log standard deviation;
pairs 3763 and 3765), so against the model's own row sums the measured ones are off by 13 and 16\,\%; holding them as
measured, in the model's units, gives correlations 0.946 and 0.997 (\texttt{device\_rows.py}).

\paragraph{Reading the Born weight itself.} The Hadamard tests read each overlap's real and imaginary parts, and the
square is taken classically. The compute--uncompute circuit of Eq.~\eqref{eq:cu} reads the Born weight
$|\braket{k_j}{q_i}|^2$ of the normalised states itself, as the probability that all four qubits return 0. It ran on
the two runs whose answer can tell the device from uniform attention, pair 3765 and pair 3763 without ``aliens''
(Table~\ref{tab:jobs}): per block one circuit per overlap and 32 calibration circuits of the same form,
compute--uncompute of random state pairs whose $|\braket{a}{b}|^2$ is known and spread evenly over $[0,1]$; each
reading $p$ becomes $(p - c)/s$ with its block's fitted slope $s$ and offset $c$, set to zero where negative (38 of
452 readings). Table~\ref{tab:hwcu} sets the two methods side by side (\texttt{cu\_analysis.py},
\texttt{device\_errors.py}, \texttt{device\_power.py}, \texttt{hw\_cu\_compare.py}); in absolute terms the calibrations
scatter by 0.024--0.027 and 0.026--0.044. A Hadamard block holds 256 overlap circuits and 32 calibrations, a
deletion block 196 and 14 shallow ones, a compute--uncompute block 128 (98 on the deletion) and 32. Unlike the visibility of the Hadamard tests,
the offset does not cancel in a row; the device's answer is the same, at 0.83--0.92, with the offset moved by
$\pm 0.02$, set to zero, or refitted on the calibrations below 0.4 alone. The slope does cancel, and enters the row
sums once, not squared ($s = 0.74$--$0.76$ against $v^2 = 0.63$--$0.66$ for the Hadamard tests on pair 3765).

\begin{table}[h]
\centering\footnotesize
\setlength{\tabcolsep}{4pt}
\begin{tabular}{lcc}
\toprule
 & Hadamard tests & compute--uncompute \\
\midrule
qubits; circuits per overlap; CZ & 5; 2 (Re, Im); 37 & 4; 1; 26 \\
quantum s, pair 3765 and the deletion & 538$^{b}$ & 316 \\
slope (visibility) $s$; readout offset $c$ & 0.67--0.81; $\le 0.006$ & 0.73--0.76; 0.02--0.03 \\
calibration scatter / shot noise & 1.15--1.33 & 2.6--4.5 \\
Born weights (3765, block 1): MAE; corr. & 0.021; 0.986 & 0.035; 0.979 \\
answer, pair 3765 (noiseless 0.905) & 0.925 (0.862--0.953) & 0.877 (0.841--0.904) \\
answer, 3763 without ``aliens'' (0.912) & 0.857 (0.762--0.913) & 0.865 (0.849--0.884) \\
row sums off per row (log s.d.) & 6--9\,\% & 11\,\% \\
evidence, 3765: corr.; $\ell_1$/sum & 0.997; 6\,\% & 0.971; 14\,\% \\
evidence, 3763: corr.; $\ell_1$/sum & 0.962; 13\,\% & not run \\
evidence, 3763 without ``aliens'' & 0.693, fails$^{a}$ & 0.48; 28\,\% \\
run & 9--10 Nov, qubits 13--73 & 12 Nov, qubits 136--144 \\
\bottomrule
\end{tabular}
\caption{The two ways of reading the attention on ibm\_kingston, from the recorded counts: 95\,\% shot-noise intervals
of the answer in brackets; MAE, mean absolute error against the ideal weights on the same vectors in both runs.
Hadamard cells cover pairs 3763 and 3765 where no pair is named. $^{a}$With its 14 shallow calibrations per block.
$^{b}$Including those calibrations. The methods ran on different days and qubits.}
\label{tab:hwcu}
\end{table}

The device gave the model's answer on both compute--uncompute runs, in every redraw, but below the noiseless model's
probability by more than shot noise: just outside the interval on pair 3765, clearly on the deletion. Its readings
are less accurate on the same quantity, the 128 Born weights of pair 3765's first block, and its calibrations scatter
2.6--4.5 times the shot noise. A quadratic fit does not lower that scatter. We did not find its cause: candidates are
the readout of four qubits at once, whose confusion depends on the whole output distribution, and the calibrations'
even spread over $[0,1]$ while the readings lie mostly low (median 0.10 in pair 3765's first block); we applied no
readout mitigation. Its blocks ran two minutes apart.

The evidence follows less closely. On pair 3765 every sign is right and the top word, the separator, is the same in
every redraw (correlation 95\,\% interval 0.942--0.987 over 500 redraws), but four of its seven words fall outside
their intervals, among them the separator (0.175 against 0.200) and the hypothesis's ``play'' (0.083 against 0.049).
Its row errors are structured: unbiased random errors of the same size give a median correlation of 0.993, below the
device's in 5 of 100 draws. On the deletion run the evidence fails (95\,\% interval 0.18--0.81): the top word is right
in 12\,\% of redraws (the device ranks the first ``men'' first, 0.131 against 0.075, and the separator, first in the
exact evidence at 0.149, second at 0.108). The exact evidence of this six-word input is flat (0.059--0.149), which
makes a correlation harsh, but the $\ell_1$ error is not a matter of the measure; random row errors of the same size
give a median of 0.975, below the device's in 1 of 100 draws. On this device and in these runs, then, the Hadamard
tests read the attention well enough for the evidence on both pairs; compute--uncompute, with one circuit per overlap
and fewer gates, reads each Born weight as a probability, up to an affine calibration, gives the right answers, and
fails the evidence on one of its two runs.

\paragraph{Resources and errors.} To see what the qubit realisation needs, we simulated it on 300 test pairs: every
overlap estimated from $S$ shots of its Hadamard test, everything else exact, as on the device
(Table~\ref{tab:resources}). The answer agrees with the noiseless model on 98--99\,\% of pairs from 1,000 shots
per circuit and on more than 99\,\% from $10^4$ (the 99\,\% mark falls between 2,000 and $10^4$ shots from one draw
of the shot noise to the next). The weight $|\braket{q_i}{k_j}|^2$ doubles an overlap's relative error (4.1\,\%
becomes 8.1\,\% at 2,000 shots), and the row normalisation removes what is common to a row. A loss of visibility
$\eta_i$ shared by the circuits of row $i$ scales each of its overlaps alike and cancels in the weights, but not in the
row sum $D_i$ that the evidence holds:
\begin{equation}
  w_{ij} = \frac{|\eta_i\braket{q_i}{k_j}|^2}{\sum_{j'}|\eta_i\braket{q_i}{k_{j'}}|^2}
         = \frac{|\braket{q_i}{k_j}|^2}{\sum_{j'}|\braket{q_i}{k_{j'}}|^2},
  \qquad
  \hat D_i = \eta_i^2 D_i .
  \label{eq:visibility}
\end{equation}
A spread of visibility from circuit to circuit does not cancel (agreement 97\,\% at a 5\,\% spread, 93\,\% at
10\,\%). A sentence of $n$ tokens takes $8(n+1)^2$ circuits, so at 2,000 shots per circuit about $10^6$ shots for
the seven-token device pairs, $9\times10^6$ for the median sentence (23 tokens) and $7\times10^7$ at 64. The hybrid
evidence is robust to shot noise (correlation with the exact evidence 0.9994 at 2,000 shots, top three words the same
in 98\,\%) and, because the row sums feel the visibility only squared (Eq.~\eqref{eq:visibility}), to the
calibration of the row sums as well: a relative error of 3\,\% in a row's visibility leaves a correlation of 0.998,
10\,\% leaves 0.968; only at 30\,\% is it lost. Squaring the weight again, as in the ablation of
Table~\ref{tab:esnli}, makes the row sums feel $\eta_i^4$: there 10\,\% leaves 0.19. How strongly an error reaches
the evidence also depends on cancellation: the ratio $\sum_t|e_g(t)|/|\sum_t e_g(t)|$ has a median of 1.09 over the
test pairs and is 1.0 on both device pairs (under the fourth power it was 8.5 on pair 3763, whose evidence there was a
small total of large opposite terms). The device runs above calibrate with circuits as deep as the overlaps, which
compile to the same gates; shallower calibrations miss loss that the overlaps suffer.

\begin{table}[!htb]
\centering\footnotesize
\setlength{\tabcolsep}{4pt}
\begin{tabular}{lcccccc}
\toprule
shots per circuit & 100 & 300 & 1,000 & 2,000 & $10^4$ & $10^5$ \\
\midrule
answer agrees & 0.87 & 0.92 & 0.987 & 0.980 & 0.993 & 1.000 \\
$|\Delta P|$, median & 0.062 & 0.034 & 0.016 & 0.013 & 0.006 & 0.002 \\
evidence: correlation & -- & -- & -- & 0.9994 & 0.9999 & 0.99999 \\
evidence: top-3 & -- & -- & -- & 0.98 & 0.98 & 0.997 \\
\midrule
visibility spread per circuit & 0 & 0.02 & 0.05 & 0.1 & 0.2 & -- \\
answer agrees ($10^4$ shots) & 0.993 & 0.997 & 0.973 & 0.933 & 0.883 & -- \\
\midrule
row-visibility error & 0.01 & 0.03 & 0.1 & 0.3 & -- & -- \\
evidence: correlation & 0.9996 & 0.998 & 0.968 & 0.00 & -- & -- \\
evidence: top-3 & 0.98 & 0.95 & 0.84 & 0.22 & -- & -- \\
\bottomrule
\end{tabular}
\caption{The Born Transformer on qubits, simulated: attention overlaps from Hadamard tests with shot noise (and, in
the lower rows, a visibility that varies by circuit, or a row-sum calibration that is off), everything else exact;
300 test pairs (evidence: 100). Correlation and top-3 agreement are against the exact evidence.}
\label{tab:resources}
\end{table}

\paragraph{How many shots the evidence needs.} The hybrid evidence $\hat e$ (a vector over the words) depends on the
device only through the measured row sums $\hat D_i = D_i(1+\delta_i)$, $i = 1, \dots, R$ over blocks, heads and
slots (each held row sum is scaled by measured over exact, as on the device), and is a smooth function of the
$\delta_i$. Its sensitivities $G_i = \partial e/\partial \log D_i$ at the
exact row sums come from the noiseless model, before any device run. With compute--uncompute readings the row sum is
linear in the counts, $\hat D_i = \sum_j n_{ij}^2(\hat p_{ij} - c)/s$, with $n_{ij} = \|q_i\|\,\|k_j\|$, $\hat p_{ij}$
the frequency of 0000 and $p_{ij} = s\,r_{ij} + c$ its expectation, $r_{ij} = |\braket{\hat k_j}{\hat q_i}|^2$; with
the slope $s$ and offset $c$ known, and without the device run's clipping at zero, it is unbiased.
\needspace{12\baselineskip}
\begin{proposition}
\label{prop:shots}
Let every compute--uncompute circuit run $S$ shots, with $s$ and $c$ fixed, and let
\[
  A_i = \frac{1}{D_i^2}\sum_j n_{ij}^4 , \qquad V_i = \frac{1}{s^2 D_i^2}\sum_j n_{ij}^4\, p_{ij}(1 - p_{ij}) .
\]
Write $\hat e - e = \Delta_1 + O(\max_i \delta_i^2)$, with the first-order error $\Delta_1 = \sum_i \delta_i G_i$.
Then:
\begin{enumerate}[label=(\alph*), leftmargin=2em, itemsep=2pt, topsep=3pt]
\item with probability at least $1 - \beta$,
\begin{equation}
  \|\Delta_1\|_1 \le \frac{1}{s}\sqrt{\frac{\ln(2R/\beta)}{2S}}\; \sum_{i=1}^{R} \|G_i\|_1 \sqrt{A_i} ;
  \label{eq:shotbound}
\end{equation}
\item in expectation,
\begin{equation}
  \mathbb{E}\,\|\Delta_1\|_2^2 = \frac{1}{S}\sum_{i=1}^{R} V_i\, \|G_i\|_2^2 .
  \label{eq:shotmean}
\end{equation}
\end{enumerate}
Hence $\|\Delta_1\|_1 \le \varepsilon\,|\sum_t e_t|$ holds with probability at least $1-\beta$ once
\begin{equation}
  S \;\ge\; \frac{\ln(2R/\beta)}{2 s^2 \varepsilon^2}\; A_{\max}\, (\gamma\rho)^2 ,
  \qquad
  \gamma = \frac{\sum_i \|G_i\|_1}{\sum_t |e_t|}, \quad \rho = \frac{\sum_t |e_t|}{|\sum_t e_t|} .
  \label{eq:shotreq}
\end{equation}
\end{proposition}
The proof (Appendix~\ref{app:shots}) is Hoeffding's inequality for each row sum, a union bound over the rows and the
first-order expansion of $\hat e$; (b) uses that different rows are read by different circuits. The bound is on the
first-order error; the full error differs from it by a remainder quadratic in the $\delta_i$. The shots grow with
the square of the cancellation $\rho$ (the ratio of the resources paragraph above) and of the
sensitivity $\gamma$. We checked the proposition on 100 random test pairs (model seed 0) with shot noise alone ($s = 1$,
$c = 0$; 40 draws per pair at 500, 2,000 and $10^4$ shots; \texttt{shot\_bound.py}). The expectation (b) predicts the
simulated root-mean-square $\ell_2$ error to between 0.78 and 1.22 times over all pairs and shot counts (median
1.00), as the sampling error of a root mean square over 40 draws leads one to expect. The bound (a) held for the 95th
percentile of the full $\ell_1$ error on every pair, at a median of 21 times above it at every shot count, and the
full error never exceeded the first-order bound $\sum_i|\delta_i|\,\|G_i\|_1$ in any draw. The bound pays for
Hoeffding's inequality, which uses the widths of the outcomes rather than their variances, for the union bound and
for adding the rows' errors with aligned signs; Eq.~\eqref{eq:shotreq}, which replaces each $A_i$ by the largest,
pays about seven times more in shots on the median pair. The first-order term accounts for the simulated error to
within 32\,\% at 2,000 shots and 15\,\% at $10^4$ (largest over pairs and draws; 67\,\% at 500). The need differs widely
between pairs: for a root-mean-square $\ell_2$ error of 5\,\% of $|\sum_t e_t|$, (b) asks for a median of 9 shots
per circuit over pairs (median $\rho = 1.11$ and $\gamma = 4.7$), 235 at the 90th percentile and 1,913 at most, and
the shots it asks for follow the cancellation (correlation 0.81 between $\log\rho$ and $\log S$). The guarantee is
far more demanding: for an $\ell_1$ error of 5\,\% with probability 0.95, (a) asks for a median of $9 \times 10^4$
shots and Eq.~\eqref{eq:shotreq} for $7 \times 10^5$, while in the simulation the $\ell_1$ error stayed within 5\,\%
in 81, 92 and 98\,\% of draws at 500, 2,000 and $10^4$ shots. The proposition therefore gives the shot-noise floor
of a device run, pair by pair from the noiseless model before it is sent, by (b), with (a) as the guarantee; it is
not a budget for today's devices. On the two compute--uncompute runs, with their fitted $s$ and $c$ and 2,000 shots,
(b) predicts root-mean-square $\ell_2$ errors of 1.1 and 1.5\,\% of $|\sum_t e_t|$, and the device gave 7.6 and
14.4\,\%; (a) allows 54 and 72\,\% in $\ell_1$, against the device's 14 and 28\,\% (\texttt{device\_shot\_floor.py}). Shot noise was not the limiting
error of the compute--uncompute runs; errors per circuit beyond it (2.6--4.5 times it in the calibrations) were, which
the proposition, with $s$ and $c$ known, does not cover. The
same sensitivities $G_i$ carry a calibration error of the row sums to the evidence. For the Hadamard tests the row sum is a sum of squared estimates, with a bias of order $1/S$ and the
visibility entering squared; we state the proposition for compute--uncompute, where the estimate is linear in the
counts.

\section{No measurement until the end: a coherent attention layer}
\label{sec:coherent}
\begin{figure}[!htbp]
  \centering
  \includegraphics[width=\linewidth]{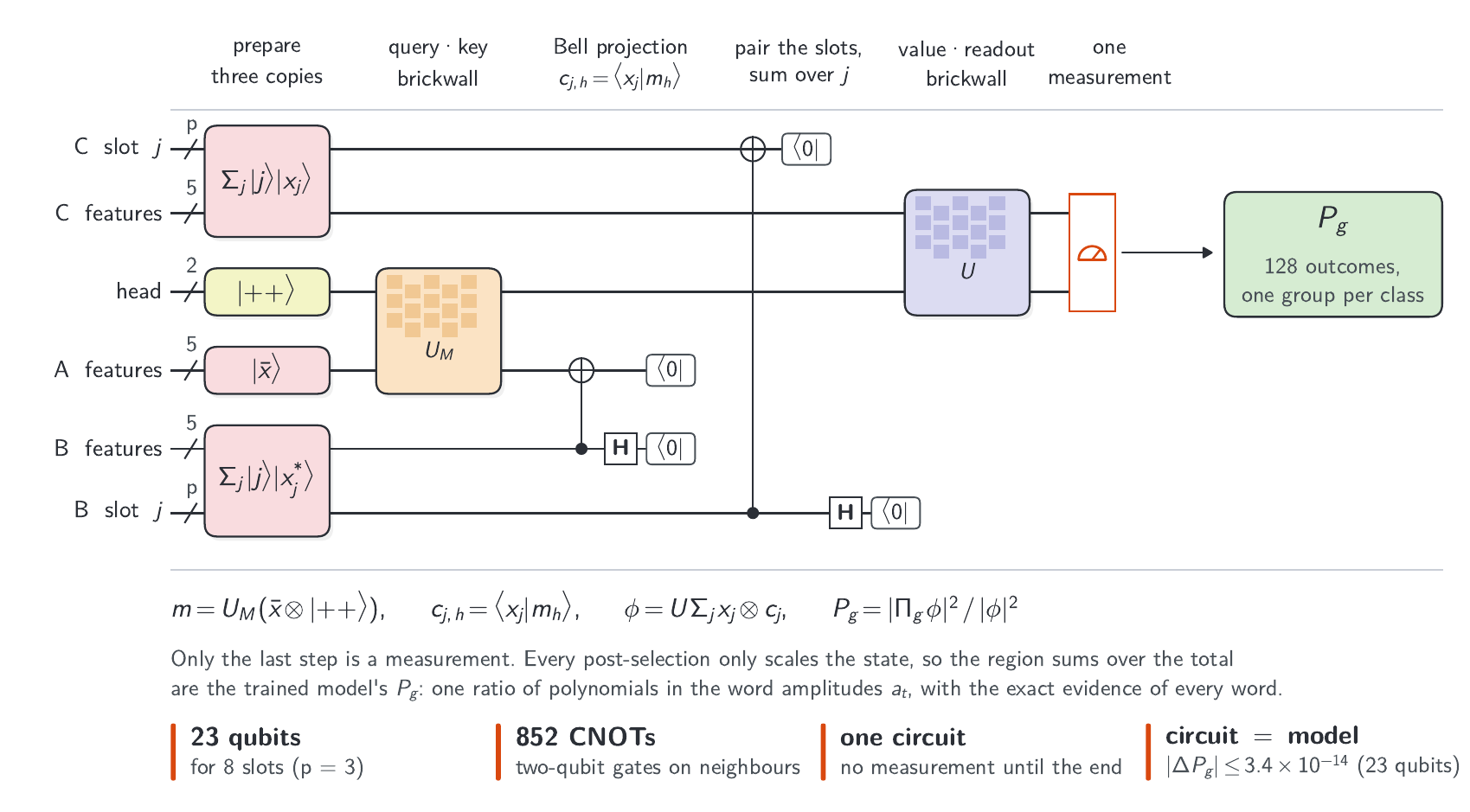}
  \caption{The coherent attention layer as one circuit, left to right: three copies of the sentence prepared,
  the query--key brickwall $U_M$, a Bell projection that leaves $c_{j,h} = \braket{x_j}{m_h}$ as an amplitude, the
  slots paired and summed, the value--readout brickwall $U$, and one measurement of seven qubits. 23 qubits, 852
  CNOTs; the circuit equals the trained model.}
  \label{fig:qubit}
\end{figure}

On qubits each reading of the Born Transformer is a separate circuit, run many times. Could attention run with no
measurement inside at all? On qubits, unlike classical light fields, copies of a state can be held together in a
tensor product, which allows it. This needs a separate, smaller model, trained on its own: it does not use the Born
Transformer's weights, and it is simulated only. We prepare three copies
of the sentence: copy A holds the pooled field $\bar x = \sum_j x_j$, copies B and C hold $\sum_j \ket{j}\ket{x_j^*}$
and $\sum_j \ket{j}\ket{x_j}$ on a slot register and five feature qubits each, and a two-qubit head register starts in
$\ket{{+}{+}}$. A map $U_M$ on (A, head) turns the pooled field into a query--key image
$m = U_M(\bar x \otimes \ket{{+}{+}})$; a Bell projection of A with B gives, for every slot and head,
\begin{equation}
  c_{j,h} = \braket{x_j}{m_h},
\end{equation}
a query--key overlap held as an amplitude. Pairing the slot registers of B and C ($j = j'$) and summing over $j$ leaves
$y = \sum_j x_j \otimes c_j$ on (C, head), which a second map $U$ (values and readout together) sends to the seven
qubits measured once: $P_g = |\Pi_g U y|^2/|U y|^2$ (Figs.~\ref{fig:compare3} and~\ref{fig:qubit}). The weights are not normalised---a
division needs a measurement---and the query is pooled over the sentence, so this is a coherent attention layer, not a
transformer. It is again a ratio of polynomials (degree 6 over one denominator), and its evidence is exact.

Trained classically with general matrices (Coherent-attn) the layer reaches $74.67 \pm 0.18\,\%$ at 20 epochs, with
exact evidence (AOPC 0.405, highlights 0.516). The circuits below are built from the 10-epoch models of
Table~\ref{tab:coherent} (seed 0; dense maps 73.14\,\%). Their equality with the models was checked on a GPU
statevector of the paired form, B and C prepared together and started from the prepared state, and for the brickwall
model also on the circuit as counted, 23 qubits with B and C apart and every state prepared by gates from
$\ket{0\cdots0}$ (to $3.4\times10^{-14}$, pass rates $2.4\times10^{-5}$ and $1.9\times10^{-4}$; \texttt{coherent23.py}). Compiled with block encodings~\cite{gilyen2019qsvt}, its circuit equals
the model to $3 \times 10^{-16}$, but needs 137k CNOTs. When every map is a brickwall of two-qubit unitaries on
neighbouring qubits---$U_M$ of 12 layers, $U$ of 24---the circuit falls to 852 CNOTs on 23 qubits, still equal to the
model, and passes post-selection about 4,900 and 1,300 times more often on test pairs 3765 and 3763, those of the Born Transformer's device runs (Table~\ref{tab:coherent}, Fig.~\ref{fig:qubit}).

\begin{table}[h]
\centering\small
\begin{tabular}{lccccc}
\toprule
coherent attention layer & params & test (\%) & CNOTs & qubits & post-selection \\
\midrule
dense maps & 82k & 73.14 & 136,727 & 27 & $10^{-9}$--$10^{-7}$ \\
brickwall readout & 18k & 72.41 & 17,452 & 25 & $10^{-8}$--$10^{-6}$ \\
\rowcolor{ours} every map brickwall (3 seeds) & 3.7k & $67.04\pm0.12$ & 852 & 23 & $10^{-5}$--$10^{-4}$ \\
\bottomrule
\end{tabular}
\caption{The coherent attention layer as one circuit (8 slots, $p = 3$; 10 epochs), on test pairs 3765 and 3763: each
circuit equals its trained model to $\le 3.3\times10^{-16}$; post-selection with the copies prepared apart. The dense
model reaches $74.67\pm0.18\,\%$ at 20 epochs over three seeds.}
\label{tab:coherent}
\end{table}

The brickwall model keeps 3.7k parameters outside the embedding and loses six points of accuracy; at 67\,\% it is
little above the linear bag of words (65.7\,\%) and the hypothesis alone (64.8\,\%, Section~\ref{sec:esnli}) and well below
the bag of embeddings (77.8\,\%), so it relates
the two sentences only weakly. Its deletion scores stay high: over three seeds (on 1,000 test pairs, at 10 epochs), deleting its top word changes the
answer as often as in the larger models (flip@1 0.48, its leave-one-out ceiling not computed; AOPC 0.35, random 0.04), and its top words agree with the
annotators' highlights at least as often (0.529; chance 0.354). Trained for 20 epochs, as the other models, it reaches $69.07 \pm 0.33\,\%$ with the
same quality of evidence (flip@1 0.46, AOPC 0.36, highlights 0.523) and the same 852-CNOT circuit, still equal to
the model ($6 \times 10^{-16}$), but passes two to three times less often ($1.2\times10^{-5}$ and $6.5\times10^{-5}$ on
the two pairs, copies prepared apart, against $2.4\times10^{-5}$ and $1.9\times10^{-4}$). The loss, a ratio of Born
probabilities, does not see the pass rate, and the sharper attention of longer training lowers it; the circuits,
tables and post-selection here use the 10-epoch models, and a penalty on the rate during training is an open
direction (Section~\ref{sec:limits}).

For the coherent attention layer the cost is the post-selection. With the copies prepared as normalised states,
the Bell projection of the five feature qubits and the sum over the $2^p$ slots each divide the amplitude, and the
probability of passing has a closed form in the quantities above:
\begin{equation}
  P_{\mathrm{pass}} = \frac{|y|^2}{|\bar x|^2 \bigl(\sum_j |x_j|^2\bigr)^2\, 2^5\, 2^p},
  \qquad y = \sum_j x_j \otimes c_j .
  \label{eq:pass}
\end{equation}
It reproduces the statevector's $2.4\times10^{-5}$ and $1.9\times10^{-4}$ on the same two seven-token pairs (3765 and 3763), and over 300 test
pairs its median falls by a factor of about six from 12--16 slots to 33--54 (Table~\ref{tab:pass}). One accepted shot takes
$1/P_{\mathrm{pass}}$ runs or, with amplitude amplification~\cite{brassard2002amplitude}, about $\pi/(4\sqrt{P_{\mathrm{pass}}})$ rounds of two
passes. This assumes reflections about the whole prepared state, the preparations included in the 852 CNOTs, and
$P_{\mathrm{pass}}$ known well enough to stop at the right round (or fixed-point amplification~\cite{yoder2014fixedpoint}).

\begin{table}[h]
\centering\footnotesize
\begin{tabular}{lccc}
\toprule
slots (words $+$ reference) & 12--16 & 17--32 & 33--54 \\
\midrule
test pairs & 28 & 224 & 48 \\
qubits & 25 & 27 & 29 \\
$P_{\mathrm{pass}}$, median & $1.6\times10^{-5}$ & $6.8\times10^{-6}$ & $2.8\times10^{-6}$ \\
runs per accepted shot, $1/P_{\mathrm{pass}}$ & $6.3\times10^{4}$ & $1.5\times10^{5}$ & $3.5\times10^{5}$ \\
amplification rounds & 200 & 300 & 470 \\
CNOTs per accepted shot, amplified & $3\times10^{5}$ & $5\times10^{5}$ & $8\times10^{5}$ \\
\bottomrule
\end{tabular}
\caption{The coherent attention layer's post-selection, Eq.~\eqref{eq:pass}, on 300 test pairs (the brickwall model,
852 CNOTs per pass, the eight-slot circuit's: a lower bound for longer sentences, whose preparations are larger). Amplification rounds: $\pi/(4\sqrt{P_{\mathrm{pass}}})$, two passes each.}
\label{tab:pass}
\end{table}

Even at 852 CNOTs, one circuit with a pass rate of $10^{-4}$ is beyond today's devices; the Born Transformer's
Hadamard tests of Section~\ref{sec:device}, five qubits each, are not.

\section{Related work}
\label{sec:related}
\paragraph{Born machines.} Born machines model data by the Born probabilities of a quantum state,
$p(x) = |\braket{x}{\psi}|^2$~\cite{cheng2018born}, prepared by a parametrised circuit and trained as generative
models~\cite{liu2018qcbm,benedetti2019generative}. Diffractive optical networks classify by the light that falls on
one detector region per class~\cite{lin2018d2nn}; our readout normalises such region intensities into the Born
probability $P_g$ and is trained on it. CNet~\cite{crasmaru2026cnet} uses the Born rule as a classification head and
as causal attention, $|\braket{Q_k}{K_j}|^2$ over its row sum, for character-level language modelling. We use the
same rule in a bidirectional classifier, as the attention and the readout, with the class probabilities explained
word by word and the model compiled to optics and qubits.

\paragraph{Quantum attention.} Quantum self-attention networks either measure query and key circuits and form the weights classically, as in Gaussian-projected attention for text classification~\cite{li2022qsann}, or measure the query--key similarity of mixed states directly by a swap test~\cite{chen2024qmsan}; quantum vision transformers add orthogonal layers and a compound-matrix attention computed
on the quantum side, run on superconducting hardware up to six qubits~\cite{cherrat2024qvit}; a survey sorts the field into circuit-based and
quantum-linear-algebra families~\cite{zhang2025survey}. Like the Born Transformer's Hadamard tests, the self-attention
networks measure between the quantum and the classical steps. Quixer prepares a superposition of token unitaries by a linear combination of
unitaries and applies a quantum singular value transform~\cite{khatri2024quixer}, closest in spirit to our coherent
layer; quantum linear algebra can block-encode a transformer layer's self-attention, residual with LayerNorm and feed-forward for fault-tolerant inference~\cite{guo2024qtransformer}, using the quantum singular value transform~\cite{gilyen2019qsvt}. A recent theoretical construction realises single-head softmax attention
with Born-rule measurements, for inputs and outputs on the probability simplex: scores read by Hadamard tests on
block-encoded projections, a $\cos^2$ readout normalised over each row that equals the softmax on the interior of the
simplex, one measure-and-reload step per attention score---hybrid, as our device runs are---and a coherent variant
that is $\varepsilon$-approximate by quantum singular value transformation in the limit of infinite depth, with a no-go result for an exact finite-depth QSVT circuit with affinely encoded scores~\cite{reinhardt2026roadmap}. It is complementary to ours: it realises the softmax itself, as a
circuit-level construction, where here the Born rule replaces the softmax in a model that is trained, explained and
run.

\paragraph{Interpretable quantum models.} QLens describes a classical transformer's layers as unitaries on a state
whose Born-rule measurement gives the output distribution, as a lens for interpreting it~\cite{gupta2025qlens}. A fully coherent quantum transformer block has been read through the mutual
information, entanglement and fidelity of its states~\cite{iacopetta2026watching}; compositional models of text
trained classically and run on trapped ions are interpretable by construction~\cite{duneau2024qdiscocirc}, the same
choice of training classically as ours; compositional circuits have also been applied to natural language
inference~\cite{sun2025qnlpnli}, and compositional models of meaning were run on IBM hardware
\cite{lorenz2023qnlp}. Complex-valued word embeddings, from the line of quantum-inspired language models, encode position in their phases~\cite{wang2020complex}, as our word fields do. Here the explanation is a decomposition of the answer itself into the words, exact with the denominators held,
and checked against people.

\paragraph{Optical transformers.} Our optics follow coherent nanophotonic meshes for neural networks
\cite{shen2017coherent} and electro-optic activations that tap light and use the detected signal to modulate it~\cite{williamson2020electrooptic}; our measured feed-forward likewise detects and re-modulates, on a fresh modulator. Transformer operations have run on an optical matrix--vector
multiplier~\cite{anderson2024optical}, and photonic accelerators target attention: its products of two dynamic
operands~\cite{zhu2024lightening} and the cost of converting signals between light and electronics
\cite{li2025hyatten}. The photonic transformer chip of Tian et al.~\cite{tian2025ptc}
replaces every multiplication of attention with an optical interference read by balanced detection and drops the
softmax, and runs a vision transformer's attention on a fabricated chip (94\,\% on 100 randomly drawn MNIST test digits). We share its two
observations---interference gives $\mathrm{Re}(a b^*)$, and attention can work without a softmax---reached there from
device physics and here from the Born rule; that work has a fabricated chip, while our optics are simulated.

\paragraph{Polynomial and bilinear networks.} Bilinear feed-forward layers, the elementwise product of two linear
images, make a network's weights readable by eigendecomposition~\cite{pearce2025bilinear}; our measured feed-forward
is of that form, read here as an interference. Polynomial attention has been used for speed~\cite{kacham2023polysketch},
and Born scores $|\braket{q_i}{k_j}|^2$ normalised by a running sum in a complex-valued transformer of 253M parameters
for byte-level language modelling, which reached each validation loss in about half the optimisation steps of its
real counterpart early in training, an advantage that decayed later, with a softmax vocabulary
head~\cite{nebli2026shortcut}. A polynomial graph network makes the Aumann--Shapley (integrated-gradient) attributions
of its pre-softmax scores exact by the same means as ours, a degree bound and Gauss--Legendre
quadrature~\cite{feng2026apex}; here the denominators are held, so that the quantity decomposed is the class
probability itself, at the input.

\paragraph{Attention and explanation.} Whether attention weights explain a model is debated
\cite{jain2019attention,wiegreffe2019attention}; our evidence is not the attention weights but an exact decomposition
of the answer. Relevance propagation through transformers conserves relevance by its rules, up to what biases and stabilisers absorb,
\cite{chefer2021transformer,achtibat2024attnlrp}; here conservation is a property of integrated gradients on a model
that is a polynomial once its denominators are held, and a finite quadrature computes it. Our deletion test is the form of the faithfulness metrics of rationale benchmarks such as
ERASER~\cite{deyoung2020eraser}, on the e-SNLI highlights~\cite{camburu2018esnli}.

\section{Limitations}
\label{sec:limits}
\begin{itemize}
  \item Trained on a GPU, not on hardware; the device reads only the trained model's attention overlaps.
  \item The real-device runs are a demonstration on the two shortest test pairs and one deletion; statistics come from
        the exact runs and the noise model.
  \item The hybrid evidence followed the exact one on both device pairs by Hadamard tests, but not on the
        compute--uncompute deletion run (correlation 0.48); only the row sums, not the numerators, come from the
        device; on the Hadamard tests the error left was shot noise and a fifth to a half more per circuit, which more
        shots per circuit reduce only in part, and a pair whose evidence cancels needs its row sums within a few per cent.
  \item The coherent circuit (852 CNOTs on 23 qubits, post-selection $10^{-5}$--$10^{-4}$) is beyond today's devices,
        above all by the noise of that depth; it equals its model on the two pairs checked, and its brickwall form costs
        six points of accuracy.
  \item The optics are simulated with shot noise (Poisson counts at every detector and camera pixel), finite
        precision of the encoded fields and weights, and mesh phase errors only: not crosstalk, phase drift of the
        coherent sums, detector efficiency and electronic noise, the resolution of the converters that read the measured
        numbers, modulator extinction and insertion loss, the camera's dark counts, read noise and dynamic range, or
        loss that differs between pixels (a uniform loss cancels in $P_g$). At $10^{10}$ source photons a balanced
        pair of the attention overlaps receives a median of $10^4$--$2\times10^5$ photons, but one of the feed-forward
        readings a median of 39 in the first block and 2.4 in the second (1\,\% of them below 0.005), so its detectors
        would need read noise well below one photon; the accuracy above holds with these counts. A sentence then carries about 1.3\,nJ at 1550\,nm, about 2\,mW per waveguide if it is
        delivered in 1\,ns. The meshes are full square ones, uncalibrated, and
        uncorrected phase errors must stay near 0.01\,rad. Each block needs at least two optical shots or an optical
        delay, and the layout of the slots in space or time is not fixed.
  \item One dataset with human highlights, and models far smaller and less accurate than the state of the art on SNLI.
  \item The guarantee of exactness grows costly with depth: the degree bound is $2\cdot10^B$ for $B$ blocks, so the
        rule needs $10^B$ nodes---100 for two blocks, $10^4$ for four. The bound is loose (30 nodes already
        agree here), but deeper models would need a tighter bound, adaptive quadrature or a lower-degree
        feed-forward.
  \item On qubits the shot cost grows with the square of the sentence length, $8(n+1)^2$ circuits per pass.
  \item The rankings compared are integrated gradients, attention, attention $\times$ gradient, gradient $\times$ input
        and leave-one-out, and, on 55 pairs of at most 14 tokens, the exact Shapley value of the deletion game
        (Appendix~\ref{sec:freeden}); relevance propagation (LRP, AttnLRP) and Shapley values on longer pairs were not.
  \item What the device computes: the Hadamard tests return the real and imaginary parts of each overlap; squaring
        them and dividing by the row sum is classical, and so is every per-word quantity and the division by the
        visibility. The compute--uncompute circuits return $|\braket{k_j}{q_i}|^2$ of normalised states itself as a
        probability, but the norms, the correction for slope and offset, the division by the row sum and every
        per-word quantity stay classical, and on this device their readings are less accurate than the Hadamard tests'.
  \item Exactness buys the sum, not the best ranking: on 55 short pairs the exact Shapley value of the deletion game
        changes the answer by its first word more often than BEW in every seed, leave-one-out does so with $n$ forward
        passes, and BEW ranks words much as gradient $\times$ input with the denominators held does. The ranking also
        depends on holding the denominators (rank correlation 0.21--0.36 with the free attribution, Appendix~\ref{sec:freeden}).
  \item The share $b_g$ is the class's share of the camera's pixels, not the model's answer with no words; the
        evidence explains the answer above that share.
  \item We claim no quantum advantage: the device runs computations a classical computer does at once.
\end{itemize}

\paragraph{Open directions.} Several of these limits are where the work could go next. On light, the model awaits a
bench or a photonic chip, with the mesh phases calibrated to within about 0.01\,rad (a short fine-tune with
simulated phase noise did not widen this margin). On qubits, the evidence has survived the device on the two shortest pairs; on many pairs, including those whose evidence
cancels, it needs row sums within a few per cent, which on the Hadamard tests means more shots per circuit and smaller errors per circuit, and on
compute--uncompute circuits smaller errors per circuit (Section~\ref{sec:device}), and
the coherent layer needs a higher pass rate---through amplitude amplification, or a penalty on the rate during
training---before it can run. Larger models, longer sentences, other tasks with human evidence, and pictures, where
Seeing Why began, are open as well. The code, the trained models and the record of every device job are released to
make these steps easier to take.

\section{Conclusion}
\label{sec:conclusion}
A transformer whose every nonlinearity is a quadratic detection, with attention in the form of the Born rule, answers
natural language inference about as well as a softmax transformer of its size and an order-free bag of embeddings in these runs, and gives each answer an exact word-by-word decomposition---exact because the model is a polynomial
once its denominators are held, physical because its numerators and held denominators are formed from detected
readings---in a way that can
be checked: against the
answer it accounts for, by deleting words, and against what people highlighted; its attention weights, as close to
what people highlighted, account for the answer less well. One set of weights compiles exactly to simulated light, and its attention to Hadamard tests or compute--uncompute circuits on qubits; in a separate, smaller model
with no measurement inside, attention becomes one post-selected circuit equal to that model in simulation. On a quantum
computer today the answer survives on the two shortest test pairs, also when the one word that decides one of them
is deleted, also with each Born weight of normalised states read as a probability. With Hadamard tests and calibration circuits as deep as the overlaps each word's evidence computed with the device's
attention row sums follows the exact one, closely on one pair and within shot noise and a modest per-circuit error on the other; with the
compute--uncompute circuits, one per overlap, it fails on one of two runs, where errors per circuit beyond shot noise set the limit. Longer pairs and evidence that cancels need row sums within a few
per cent, and the coherent layer a higher pass rate. Keeping the polynomial's degree under control through attention and two blocks is what makes a finite quadrature
exact. We hope the evidence, and not accuracy alone, becomes part of how such models are judged.

\section*{Acknowledgements}
We acknowledge the use of IBM Quantum services for this work. The views expressed are those of the authors, and do
not reflect the official policy or position of IBM or the IBM Quantum team.

\section*{Code and data availability}
The code, the trained models of the tables (the four models of Table~\ref{tab:esnli}, three seeds each, and the coherent attention layer), an interactive notebook that explains any premise and
hypothesis and lets the reader delete words (runs on a CPU in Colab), a training notebook, and the ledger of every
hardware job (circuits, transpiled circuits, calibrations and raw counts) are at
\url{https://github.com/stein-research/reading-why}, archived with every version at Zenodo,
\href{https://doi.org/10.5281/zenodo.23938333}{doi:10.5281/zenodo.23938333}.

\section*{Author contributions and use of AI tools}
T.T. conceived the study, developed the methods and the code, ran the simulations and the hardware experiments, and
wrote the manuscript. J.P. and R.J. worked with T.T. on the physics and the optics of the model, reviewed the
manuscript and advised on its content and presentation. An AI assistant (Claude, Anthropic) was used under the
authors' direction to write and run simulation and analysis code, to prepare and check the hardware jobs, to write
the code that draws the figures, to search the literature and check the references against their sources, and to
draft and revise the text; the authors chose what to cite. The authors retained responsibility for the study design,
the scientific judgments and the claims, checked the reported results against the saved data and the code, approved
the final version, and take full responsibility for its content.

\clearpage
\appendix
\section{Model, training and hardware details}
\label{app:details}

Table~\ref{tab:models} sets the four models side by side, Table~\ref{tab:training} their training and evidence,
and Table~\ref{tab:jobs} the device runs.

\begin{table}[H]
\centering\footnotesize
\setlength{\tabcolsep}{3pt}
\renewcommand{\arraystretch}{1.15}
\begin{tabular}{>{\raggedright\arraybackslash}p{0.15\linewidth}*{4}{>{\raggedright\arraybackslash}p{0.195\linewidth}}}
\toprule
 & \textbf{Born Transformer} & \textbf{softmax} & \textbf{Poly-sim-born} & \textbf{coherent attention layer} \\
\midrule
state & complex, 32 modes & real, $d = 64$ & real, $d = 64$ & 3 copies of 5 feature qubits, a slot register of $p = 3$ qubits on two of them, 2 head qubits \\
blocks, heads & 2, two of 16 modes & 2, two & 2, two & 1, four (a 2-qubit register) \\
attention & $|\langle q|k\rangle|^2$ over its row sum & softmax & $(q\cdot k)^2$ over its row sum & Bell projection $\langle x_j|m_h\rangle$, unnormalised \\
feed-forward & $\mathrm{Re}[(W_1h)\overline{(W_2h)}]$, 170 channels, gain, $W_3$ & GELU, width 256 & bilinear, width 170 & none \\
normalisation & none & LayerNorm & none & none \\
positions & additive fields, a phase per position & learned, additive & learned, additive & additive fields, a phase per position, a field per sentence \\
readout & camera of 256 pixels, three regions, $P_g$ & mean pooling, linear & mean pooling, linear & 7 qubits read, 128 outcomes in three groups \\
maps & dense & dense & dense & brickwalls: 12 layers ($U_M$), 24 ($U$) \\
parameters & 1,530,593 (98,081) & 1,532,803 (100,291) & 1,530,752 (98,240) & 1,436,192 (3,680) \\
evidence & exact, 100 nodes & IG, 256 nodes & exact, 50 nodes & exact, 3 nodes \\
\bottomrule
\end{tabular}
\caption{The models. Parameters: total and, in brackets, outside the word embedding. The Born Transformer also
has a reference slot with a learned gain; its embeddings and position fields start at scale 1.}
\label{tab:models}
\end{table}

\begin{table}[H]
\centering\small
\begin{tabular}{>{\raggedright\arraybackslash}p{0.17\linewidth}>{\raggedright\arraybackslash}p{0.76\linewidth}}
\toprule
data & e-SNLI: 549,367 training, 9,842 dev, 9,824 test pairs; premise \texttt{<sep>} hypothesis, at most 64 tokens; vocabulary of 22,319 entries (22,316 words seen at least 3 times, and the special tokens) \\
loss & cross-entropy of $P_g$ (Born Transformer, coherent layer), of the logits (softmax, Poly-sim-born) \\
optimiser & Adam, learning rate $2\times10^{-3}$ (chosen on dev for the softmax model and the fourth-power Born model, and used untuned for the others, the Born Transformer included), cosine decay, gradient norm clipped at 1 \\
schedule & batch 64; 20 epochs (coherent layer: 10 in the tables, 20 also reported); word dropout 0.1; seeds 0--2; the last epoch is reported \\
precision & float32 for training, float64 for evidence and compilation \\
evidence & integrated gradients from the zero baseline with the denominators held; Gauss--Legendre with half the degree in nodes (Born Transformer: degree 200, 100 nodes) \\
\bottomrule
\end{tabular}
\caption{Training and evidence, the same for every model unless stated.}
\label{tab:training}
\end{table}

\begin{table}[H]
\centering\small
\begin{tabular}{>{\raggedright\arraybackslash}p{0.43\linewidth}lcc}
\toprule
job & id & circuits & quantum s \\
\midrule
pair 3763, block 1 & dbormf5t7fqc7388atrg & 288 & 155 \\
pair 3763, block 2 & dbos24s4148s739tgu50 & 288 & 155 \\
pair 3765, block 1 & dbovg2s4148s739tmj90 & 288 & 155 \\
pair 3765, block 2 & dbp6iltt7fqc7388rmgg & 288 & 155 \\
pair 3763 without ``aliens'', block 1 & dbp2rvgqtr0c738dm8cg & 210 & 114 \\
pair 3763 without ``aliens'', block 2 & dbp6gu5t7fqc7388rjc0 & 210 & 114 \\
\midrule
\multicolumn{4}{l}{compute--uncompute} \\
pair 3765, block 1 & dbqkh87v5hds73dg7k7g & 160 & 87 \\
pair 3765, block 2 & dbqki99abdes73foe1b0 & 160 & 87 \\
pair 3763 without ``aliens'', block 1 & dbqkjlvv5hds73dg7q10 & 130 & 71 \\
pair 3763 without ``aliens'', block 2 & dbqkklnv5hds73dg7ra0 & 130 & 71 \\
\bottomrule
\end{tabular}
\caption{Jobs on ibm\_kingston (Heron r2): SamplerV2, 2,000 shots, optimisation level 3, the best of 8 transpiler
seeds, dynamical decoupling (XY4); Qiskit 2.6.1, qiskit-ibm-runtime 0.51.0. What each block holds: Section~\ref{sec:device}.
Circuits, calibrations and raw counts are in the released ledger; the fourth-power model's jobs (versions 1--2) are in
its history.}
\label{tab:jobs}
\end{table}

\paragraph{Squaring the attention weight again.} The weight $|\braket{q_i}{k_j}|^2$ over its row sum is flatter than a
softmax: in one-block models of an earlier stage of this work (synthetic word-order task, Seeing Why's headroom
readout), the largest weight a token gave was on average about 0.29, against 0.84 under the softmax, and squaring it
again, $|\braket{q_i}{k_j}|^4$ over its row sum, raised that to about 0.50 (two seeds) with accuracy within the spread of three seeds. On a synthetic
task of recalling a value by its key among eight pairs, two-block models of the form used here learned it on every
seed with the fourth power, while with the square one seed of three stalled. On e-SNLI the two are level on accuracy and deletion, the square 0.01 higher in agreement
(Table~\ref{tab:esnli}), the square at under a third of the degree (200 against 648) and with row sums that feel a loss of
visibility squared rather than to the fourth power (Section~\ref{sec:device}); the paper's model is the square. Versions 1--2 of this paper used the fourth power as
the Born Transformer; this version uses the square for its lower degree and its row sums' milder dependence on
visibility. On dev, too, the square leads the fourth power (77.55 against 77.30\,\%; test 77.45 against 76.98\,\%), a
difference within the spread of the seeds: the square was chosen, after the fourth-power model's device runs, for the
two reasons above, not for this difference. The learning rate was chosen on dev
for the softmax model and the fourth-power model and used for the square without retuning, on one seed at 10 epochs
from four values for the softmax model ($5 \times 10^{-4}$ to $5 \times 10^{-3}$; dev 70.31, 74.22, 74.32 and
63.37\,\%) and three for the fourth-power model ($5 \times 10^{-3}$ diverged), so
$10^{-3}$ and $2 \times 10^{-3}$ differ by 0.1 point; no warm-up or weight decay was tried, and the softmax baseline is
lightly tuned. At the last epoch the
softmax model fits its training set less closely than the Born Transformer (training accuracy 82\,\% against
85--87\,\%), so a longer schedule might still raise it. The headline numbers and
the device runs are new. The fourth-power model ran on the same device first, and every outcome is in the released
ledger's history: its answers matched the noiseless model on all four pairs it ran (3763, 3765, 997 and 4667) and
under the deletions asked; its evidence with the measured row sums correlated with the exact evidence at $-0.86$ and
0.48 on pairs 3763 and 3765, whose calibrations were shallower than the overlaps, and at 0.93 and 0.997 on pairs 997
and 4667, with calibrations as deep as the overlaps. That difference is why the runs reported here use depth-matched
calibrations, and the fourth power's row sums, which feel a loss of visibility to the fourth power, are one of the two
reasons for the square.

Everything except the device runs was done on one laptop (Table~\ref{tab:compute}).

\begin{table}[H]
\centering\scriptsize
\begin{tabular}{>{\raggedright\arraybackslash}p{0.34\linewidth}>{\raggedright\arraybackslash}p{0.60\linewidth}}
\toprule
\multicolumn{2}{l}{\emph{hardware}} \\
GPU, CPU, memory & NVIDIA GeForce RTX 5050 Laptop GPU (8\,GB); Intel Core i5-13450HX; 16\,GB \\
quantum computer & IBM ibm\_kingston (Heron r2) through Qiskit Runtime \\
\midrule
\multicolumn{2}{l}{\emph{software}} \\
Python 3.12.3 & \\
PyTorch 2.11 (CUDA 12.8)~\cite{paszke2019pytorch} & models, training, evidence, optical compile \\
NumPy 2.5, SciPy 1.18 & arrays, data \\
\texttt{interferometer} 1.1.2 & outside check of the optical build (Clements meshes~\cite{clements2016optimal}) \\
Qiskit 2.6.1~\cite{javadi2024qiskit} & circuits, transpiling \\
qiskit-aer 0.18.2 & noisy simulations (FakeKingston) \\
qiskit-aer-gpu 0.18.2, cuQuantum 24.8 & statevectors of the coherent attention layer (cuStateVec) \\
qiskit-ibm-runtime 0.51.0 & device jobs \\
Matplotlib 3.11 & figures \\
\midrule
\multicolumn{2}{l}{\emph{time}} \\
training, 20 epochs (seed 0) & 69\,min Born Transformer; 18\,min softmax; 32\,min Poly-sim-born \\
coherent attention layer, 10 epochs & 28\,min \\
ibm\_kingston & 1,164 quantum seconds (Hadamard tests: pairs 3763 and 3765, 310 each, the deletion, 228; compute--uncompute: 316) \\
\bottomrule
\end{tabular}
\caption{Computational setup.}
\label{tab:compute}
\end{table}

\section{Explaining with the denominators free}
\label{sec:freeden}

\begin{figure}[H]
  \centering
  \includegraphics[width=\linewidth]{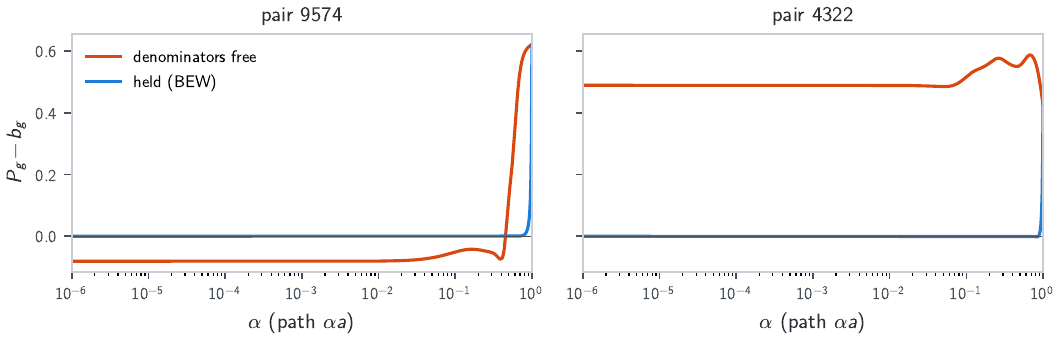}
  \caption{$P_g - b_g$ along the path $\alpha a$, with the denominators held (BEW) and free, on two test pairs
  (seed 0).}
  \label{fig:freepath}
\end{figure}

Integrated gradients can also be taken along the same path with every denominator recomputed at each point. We did
so for the same three Born Transformers (seeds 0--2) on the 1,000 test pairs of Table~\ref{tab:esnli}, with 1,024
nodes. The free attributions are complete in their own terms: they add up, to a median of $7\times10^{-15}$ (largest
$2\times10^{-3}$, one pair), to $P_g(a)$ minus the limit of the free path at zero. That limit is not $b_g$. The reference slot is
divided by $\sum_t a_t$, which is held for BEW but recomputed here, so it does not shrink with the amplitudes; as they
vanish, the model still gives an answer, and the free path starts there. Fig.~\ref{fig:freepath} shows it flat below
$\alpha \approx 10^{-2}$, at a median distance from $b_g$ of 0.23--0.30 over seeds (largest 0.65). The free
attributions therefore decompose the answer above the model's answer for vanishing amplitudes, which carries no
meaning for the words; held, the path is a polynomial that starts at zero, and its total is $P_g - b_g$. (Our code
adds $10^{-30}$ to every denominator against a division by zero; with free denominators the row sums fall below it
near $\alpha \approx 10^{-7}$, so these runs use $10^{-300}$.)

The free ranking differs from BEW's (median rank correlation 0.21--0.36 over seeds; the same top word in 25--49\,\%
of pairs) and accounts for deletion less well: deleting its top word changes the answer in 33\,\% of pairs against
44\,\% for BEW, and its five lowest in 6\,\% against 1\,\%. On the 55 pairs of at most 14 tokens, the exact Shapley
value of the deletion game itself, every denominator recomputed for each of the $2^n$ subsets of words and the empty
set at $b_g$, ranks the words closer to BEW than to the free integrated gradients in two seeds and as close in the
third (median rank correlation 0.44--0.50 against 0.22--0.44; the same top word as BEW in 44--45\,\%). Its flip@1 is
higher than BEW's on the same pairs in every seed (0.55, 0.44 and 0.38 against 0.33, 0.29 and 0.33): the deletion game
itself accounts for deletion best, at the cost of $2^n$ evaluations of the model (16,384 for the longest of these
pairs), where BEW takes 100 gradient evaluations at any length for this two-block model (the cost of each grows with
the length).

\section{Proof of Proposition~\ref{prop:shots}}
\label{app:shots}
Fix a pair and a class.

\paragraph{Each row sum.} A row sum's relative error is a weighted sum of independent shot outcomes,
\[
  \delta_i = \frac{\hat D_i}{D_i} - 1 = \frac{1}{s D_i}\sum_j n_{ij}^2\,(\hat p_{ij} - p_{ij})
           = \sum_j \sum_{k=1}^{S} \frac{n_{ij}^2}{s S D_i}\,(b_{ijk} - p_{ij}),
\]
with $b_{ijk} \in \{0, 1\}$ the outcome of shot $k$ of circuit $(i, j)$ (1 for 0000). The terms are independent,
have mean zero, and each lies in an interval of width $n_{ij}^2/(sSD_i)$, so Hoeffding's inequality gives
\[
  \Pr\bigl(|\delta_i| \ge \tau\bigr)
  \le 2\exp\!\Bigl(-\frac{2\tau^2}{\sum_j S\,\bigl(n_{ij}^2/(sSD_i)\bigr)^2}\Bigr)
  = 2\exp\!\Bigl(-\frac{2Ss^2\tau^2}{A_i}\Bigr).
\]

\paragraph{All rows.} Setting the right-hand side to $\beta/R$ and taking the union over the $R$ rows, with
probability at least $1 - \beta$ every row has
\[
  |\delta_i| \le \frac{1}{s}\sqrt{\frac{A_i \ln(2R/\beta)}{2S}} .
\]

\paragraph{The evidence, part (a).} The hybrid evidence is a smooth function of
$\log \hat D_i = \log D_i + \log(1+\delta_i)$, so
\[
  \hat e - e = \sum_i \delta_i G_i + O\bigl(\max_i \delta_i^2\bigr),
  \qquad
  \|\Delta_1\|_1 \le \sum_i |\delta_i|\,\|G_i\|_1 ,
\]
which with the bound on each $|\delta_i|$ gives Eq.~\eqref{eq:shotbound}. Since
\[
  \sum_i \|G_i\|_1\sqrt{A_i} \le \sqrt{A_{\max}}\,\gamma \sum_t|e_t| = \sqrt{A_{\max}}\,\gamma\rho\,\Bigl|\sum_t e_t\Bigr| ,
\]
the right-hand side of Eq.~\eqref{eq:shotbound} is at most $\varepsilon|\sum_t e_t|$ under Eq.~\eqref{eq:shotreq}.

\paragraph{Part (b).} The rows are read by disjoint sets of circuits and $s$, $c$ are fixed, not estimated, so the
$\delta_i$ are independent with mean zero and
\[
  \mathrm{Var}\,\delta_i = \frac{1}{S s^2 D_i^2}\sum_j n_{ij}^4\,p_{ij}(1-p_{ij}) = \frac{V_i}{S},
\]
hence
\[
  \mathbb{E}\Bigl\|\sum_i \delta_i G_i\Bigr\|_2^2 = \sum_i \mathrm{Var}(\delta_i)\,\|G_i\|_2^2 = \frac{1}{S}\sum_i V_i\,\|G_i\|_2^2 .
\]

The expansion is first order: the remainder is quadratic in the $\delta_i$, of relative size $O(S^{-1/2})$; in the
check of Section~\ref{sec:device} it stays below 33\,\% of the error from 2,000 shots.

\bibliographystyle{quantum}
{\raggedright
\bibliography{refs}
}

\end{document}